\documentclass[a4paper]{JHEP3}
\usepackage{epsfig,multicol,bbm}
\usepackage{amsmath}
\usepackage{amssymb}
\usepackage{amsfonts}
\usepackage{graphicx,epsfig}
\usepackage{subfigure}
\usepackage{exscale}
\usepackage{float}
\usepackage[numbers,sort&compress]{natbib}
\usepackage{bm}% bold math

\title{\boldmath Boundary action and profile of effective bosonic strings}

\author{A. S.\ Bakry\\
Institute of Modern Physics, Chinese Academy of Sciences, Gansu 730000, China\\	
E-mail:\email{ahmed.bakry@mail.com}}

\author{M. A.\ Deliyergiyev\\
Department of Nuclear and Particle Physics, University of Geneva, CH-211, Switzerland\\
E-mail:\email{maksym.deliyergiyev@unige.edu.ch}}
\author {A. A.\ Galal\\
Department of Physics,  Al Azhar University, Cairo 11651, Egypt \\
}

\author{M. N.\ Khalil\\
Department of Mathematics, Bergische Universit\"at Wuppertal, 42097 Germany\\
Department of Physics, University of Ferrara, Ferrara 44121, Italy\\
Computation-based Science Research Center, Cyprus Institute, Nicosia 2121, Cyprus.\\
Department of Physics,  Al Azhar University, Cairo 11651, Egypt \\
}

\abstract{The mean-square width of the energy profile of the bosonic string is calculated considering two boundary terms in the effective LW action. The perturbative expansion of the Lorentz-invariant boundary terms at the second and the fourth order in the effective action is taken around the free Nambu-Goto string. The calculation is presented for open strings with Dirichlet boundary condition on a cylinder.}

\preprint{}

\begin{document}
%\maketitle
%\flushbottom

%%%%%%%%%%%%%%%%%%%%%%%%%%%%%%%%%%%%%%%
%%%%%%%%%%%%%%%%%%%%%%%%%%%%%%%%%%%%%%%
\section{Introduction}

  The confinement of quarks is an essential property of quantum chromodynamics (QCD) and strong interactions. Despite the tremendous research efforts to provide mechanisms of quark confinement based dynamical gluonic degrees of freedom of the QCD, there is no compelling analytical construction for the phenomenon of confinement starting from basic principles.

The confinement property can be probed directly in the Monte-Carlo evaluations of QCD path integrals. The computer simulations of the infinitely heavy quark-antiquark ($Q\bar{Q}$) bound state revealed the linear rise property~\cite{Creutz:1980zw,Creutz:1983ev,Creutz:1980hb,Creutz:1980vq} of confining potential.

  It is understood from this observation that the linear rise of the potential at long separation between two static $Q\bar{Q}$ is due to the gluonic field~\cite{Fukugita:1983du, Flower:1985gs, Wosiek:1987kx, Sommer:1987uz, DiGiacomo:1989yp, DiGiacomo:1990hc, Bali:1994de, Haymaker:1994fm, Cea:1995zt,Okiharu:2003vt} which appears to be condensed into a stringlike flux tube. The formation of stringlike flux tube in between the color sources provides a prospective technique for the confinement of quarks. In this model, the nonperturbave properties of the QCD flux-tubes are expected to conform with that of an effective bosonic strings with fixed edges in the long string limit. 

  The string formation manifests in many strongly correlated systems ~\cite{PhysRevB.78.024510,2007arXiv0709.1042K,Nielsen197345,Lo:2005xt} and can be described after roughening transition by an effective string action. The effective string action is a low energy effective field theory~\cite{PhysRev.177.2247} which is obtained by integrating out the degrees of freedom of Yang-Mills (YM) vacuum in the presence of two static quarks. This string description supplies a tool to predict a set of infrared (IR) observables ~\cite{Caselle:2015tza,GIDDINGS198955,Bali:2013fpo,Kalashnikova:2002zz,Grach:2008ij,GIDDINGS198955,Caselle:2013qpa,Johnson:2000qz,Kalaydzhyan:2014tfa,Makeenko:2010dq,Dubovsky:2012sh} which can be confronted with the outcomes from the numerical lattice data.

  The L\"uscher term is an essential ~\cite{Jaimungal:1998hk,Caselle:2011vk} prediction associated with the string's quantum fluctuations at zero temperature. This Casimir energy is a detectable subleading correction to the linearly rising $Q\bar{Q}$ potential~\cite{Alvarez:1981kc,Arvis:1983fp} which is universal to any gauge group. The L\"uscher term has been detected in the numerical simulations of either Wilson~\cite{Billo:2011fd} or Polyakov loop correlation function representing  $Q\bar{Q}$ static pair~\cite{Caselle:818185,Juge:2002br,HariDass2008273,Caselle:2016mqu,caselle-2002,Pennanen:1997qm,Brandt:2016xsp,Gliozzi-1994}. The Y-string~\cite{Jahn2004700,deForcrand:2005vv,Pfeuffer} binding the three quark ~\cite{Bakry:2016aod,Bakry:2014gea} in baryonic configurations produces a counterpart L\"uscher-like term which have been detected in abelian lattice gauge system~\cite{deForcrand:2005vv}. 
 %%%%%%%%%%%%%%%%%%%%%%%%%%%%%%%%%%%%%%%%%%%%%%
  
  Not only the static potential but also the energy chart of the QCD vacuum in the presence of the confined color sources~\cite{Battelli:2019lkz} and its characteristic broadening is a fundamental source to probe the physics of the confinement from first principles. Many lattice simulations of many gauge groups have unambiguously verified~\cite{Caselle:1995fh,Bonati:2011nt,HASENBUSCH1994124,Caselle:2006dv,Bringoltz:2008nd,Athenodorou:2008cj,Juge:2002br,HariDass:2006pq,Giudice:2006hw,Luscher:2004ib,Pepe:2010na} the predicted logarithmic broadening~\cite{Luscher:1980iy} of the confining strings and the linear behavior~\cite{Caselle:2010zs} near the confinement point and large distances.

   Nevertheless, the analysis of string's fine structure in the lattice numerical data for the broadening profile revealed  substantial deviations~\cite{PhysRevD.82.094503, Bakry:2010sp, Bicudo:2017uyy} from the free-string Nambu-Goto (NG) model in the intermediate distance region at high temperatures. The excited spectrum~\cite{Luscher:2004ib} and finite temperature string tension ~\cite{Pisarski,PhysRevD.85.077501, Kac} obtained from the partition function of the leading-order Gaussian formulation of NG string action show a similar disagreement with the numerical data ~\cite{Juge:2002br, Kac, Bakry:2010sp, Bakry:2017fii,Bakry:2020ebo,Bakry:2020flt,Bakry:2020akg} for distance scales less than $1.0$ fm. 
     
   The higher-order corrections are terms of the asymptotic expansion in the inverse powers of color source separation distance $1/R$ associated with the interaction terms of growing dimensions. The next to leading correction terms of the NG action~\cite{PhysRevD.27.2944, Aharony:2010cx,Billo:2012da} have been subjected to numerical investigations~\cite{Caselle:2004er,Caselle:2005xy,Caselle:1994df,Caselle:2006dv,Caselle:2004jq,Pepe:2010na,Gliozzi:2010zv,Athenodorou:2010cs,Athenodorou:2011rx} to ascertain the relevance order by order to the discrepancy from the numerical data in each gauge model. In fact, there is no proof neither evidence that all orders~\cite{Alvarez:1981kc, Arvis:1983fp} of the power expansion are universal.

   The effective bosonic string theory serves as a functional tool in many QCD processes ~\cite{Caselle:2015tza,GIDDINGS198955,Bali:2013fpo,Kalashnikova:2002zz,Grach:2008ij,GIDDINGS198955,Caselle:2013qpa,Johnson:2000qz,Kalaydzhyan:2014tfa,Makeenko:2010dq,Dubovsky:2012sh}. The precise study of the strings effects suggests considering other features beyond the free NG action~\cite{Ambjorn:2014rwa}, in particular, possible stiffness properties~\cite{POLYAKOV19, Kleinert:1986bk} and boundary term corrections~\cite{Caselle:2014eka,Brandt:2017yzw,Aharony:2010db,Hellerman:2016hnf}.

   The Lorentz invariant boundary corrections~\cite{Aharony:2010db} to the static $Q\bar{Q}$ potential have shown viable in both the Wilson and the Polyakov-loop correlators cases~\cite{Caselle:2014eka,Brandt:2017yzw, Bakry:2017fii,Bakry:2020ebo,Bakry:2020flt}. The boundary corrections to the static quark potential provide an explanation to the deviations lately discovered among predictions of the effective string and numerical outcomes~\cite{Billo:2011fd}.
    
   The broadening profile of the energy field should receive similar corrections from the Lorentz invariant boundary terms in the action. However, the contributions of the boundary action to the width profile have neither been theoretically calculated nor confronted with the numerical lattice data. These corrections are hoped to account to many features of the fine structure of the profile of QCD flux-tube in IR region. In particular it could account for the well known deviations at relatively short distances at low temperatures, or larger distances for the excited spectrum of the flux-tubes~\cite{Brandt:2017yzw,Brandt:2010bw} and at high temperatures~\cite{Bakry:2018kpn,Bakry:2017fii,Bakry:2020ebo,Bakry:2020flt}. 
    
   The goal of the present paper is to analytically estimate the mean-square width resulting from the boundary terms in L\"uscher-Weisz (LW) effective string action. The calculations are laid out for open string with Dirichlet boundary condition on a cylinder. This could be compatible with the energy fields set up by a static mesonic configurations. We consider the perturbative expansion of two boundary terms at the order of fourth and six derivative and evaluate the modification to the mean-square width around the free NG string.
   
%%%%%%%%%%%%%%%%%%%%%%%%%%%%%%%%%%%%%%%%%%%%%%
%%%%%%%%%%%%%%%%%%%%%%%%%%%%%%%%%%%%%%%%%%%%%%

\section{Effective action of bosonic strings}   
  Long stringlike objects are not uncommon in field theory. The magnetic vortices in superconductors~\cite{DiGiacomo:1999a, DiGiacomo:1999b}, cosmic strings~\cite{Bukenov:1993im, Hindmarsh:2014rka} admits an effective string descriptions as well. The property of linearly rising potential suggests that the YM vacuum admits the presence of an object such as a quite thin long string, which is responsible for transmitting the strongly interacting forces among the quarks. The conjecture is in consistency~\cite{Olesen:1985pv} with the dual superconductive~\cite{Mandelstam76, Bali1996, DiGiacomo:1999a, DiGiacomo:1999b, Carmona:2001ja, Caselle:2016mqu} QCD vacuum and the formation of a vortex line dual to the Abrikosov by the virtue of the dual Meissner effect. 

   The effective field theory description holds at distance scales larger than the scale intrinsic thickness of the string~\cite{Vyas:2010wg,Caselle:2012rp}. The classical long string solution entails, however, the breaking of the $(D-2)$ transverse translational symmetry leading to the associated transverse oscillations of massless Goldston modes~\cite{GODDARD1973109,Low:2001bw}. 
   
    The Lagrangian of the low energy effective field theory~\cite{Brandt:2016xsp} may be constructed from all the terms respecting the imposed symmetries on  the system~\cite{Cohn:1992nu}. To constrain the effective action of the confining string Polchinski-Strominger (PS)~\cite{PhysRevLett.67.1681} introduced a conformal gauge on the worldsheet. Within this formalism, the action is constraint by requiring the effective field theory to be ghost free which fixes the central charge to be D=26~\cite{PhysRevLett.67.1681,Dubovsky:2012sh}. The conformal theory is manifestly Lorentz-invariant and the leading term of PS action coincides with the 4-derivative term of the NG string~\cite{Dubovsky:2012sh}. However, it appears that this technique is untraceable especially at higher orders.
    
  However, L\"uscher, Symanzik and Weisz~\cite{Luscherfr,Luscher:2002qv} suggested that the effective action of the string connecting a stable pair of $Q\bar{Q}$ in D-dimensional confining the theory of Yang-Mills can be constructed by the leading term of NG, free, action. In fact, the LW effective action admits all terms which preserves Lorentz and transnational-invariance~\cite{Meyer:2006qx} and is expressed in physical degrees of freedom. The effective action respects Lorentz symmetry through nonlinear realization of this symmetry~\cite{ISHAM197198,Gliozzi:2011hj,Aharony:2011gb} since the worldsheet gauge diffeomorphism is fixed to static/physical gauge. The nonlinearly-realized Lorentz symmetry constrains the coefficients of the higher-order terms in the action which is found to coincide with that of the NG action up to $\mathcal{O}(1/R^3)$ in the long string expansion.    

 As mentioned above, an effective string action can be constructed from the derivative expansion of collective string co-ordinates fulfilling Poincare and parity invariance. The operators of the Goldstone fields in the action are perturbations around the classical solution and  necessarily are space-time derivatives to preserve the translation invariance. In addition, the terms in the action that are proportional to the equation of motion or its derivatives do not contribute in perturbation theory and can be absorbed by field redefinition. These fluctuations $X^i$ (in $D-2$ space-time dimensions) are massless and in the static/physical gauge $X^{1}=\zeta_{1}, X^{4}=\zeta_{2} $ are restricted to transverse directions  ${\cal C}$. With this prescription, the LW effective action ~\cite{Luscherfr,Luscher:2002qv} up to 4-derivative term come into the form
%---------------------------------------------   
\begin{equation}
\begin{split}
  S_{LW}[\mathbf{X}]&=S_{cl}+\dfrac{\sigma}{2} \int d^{2} \zeta \Big(\dfrac{\partial \mathbf{X}}{\partial \zeta_{\alpha}} \cdot \dfrac{\partial \mathbf{X}}{\partial \zeta_{\alpha}}\Big) +\sigma \int d^{2} \zeta  \Bigg[c_2 \Big( \dfrac{\partial \mathbf{X}}{\partial \zeta_{\alpha}} \cdot \dfrac{\partial \mathbf{X}}{\partial \zeta_{\alpha}} \Big)^2 + c_3 \Big( \dfrac{\partial \mathbf{X}}{\partial \zeta_{\alpha}} \cdot \dfrac{\partial \mathbf{X}}{\partial \zeta_{\beta}} \Big) ^2 \Bigg]\\
  &+\gamma \int d\zeta^2 \sqrt{g} \mathcal{R}+ \alpha\int d\zeta^2 \sqrt{g} \mathcal{K}^2+...+S_{b},
\label{LWaction}
\end{split}
\end{equation}
%---------------------------------------------
where we have considered the Euclidean signature. The vector $X^{\mu}(\zeta^{0},\zeta^{1})$ maps the area $ C \subset \mathbb{R}^{2} $ into $\mathbb{R}^{4}$, the geometrical invariants $\mathcal{R}$ and $\mathcal{K}$ define Ricii-scalar and the extrinsic curvature ~\cite{POLYAKOV19, Kleinert:1986bk} of the corresponding configuration of world sheet, respectively. The term $S_{cl}$ characterizes the classical term, on the quantum level the Weyl invariance of the action is broken in four dimensions; however, the anomaly is known to vanish at large distances \cite{Olesen:1985pv}.

Cylindrical boundary conditions explicitly break translation invariance of the classical solution, this would entail the emergence of terms other than that in bulk of the effective action. The boundary action $S_{b}$ is a surface term located at the boundaries $\zeta^{1}=0$ and $\zeta^{1}=R$. This ought to signifies the interplay of the effective string with either the Polyakov loops on the fixed ends of the string or Wilson's loop which we will discuss in detail in the next section. 

The couplings $c_1$, $c_2$ in LW action Eq.~\eqref{LWaction} are the parameters of effective low-energy theory. For the next-to-leading order terms in $D$ dimension the open-closed duality~\cite{Luscher:2004ib} imposes further constraint on these kinematically-dependent couplings to
%---------------------------------------------
\begin{equation}
(D-2) c_2+c_3=\left( \dfrac{D-4}{2\sigma} \right).
\label{couplings1}
\end{equation}
%---------------------------------------------
Moreover, it has been shown~\cite{Billo:2012da} through a nonlinear Lorentz-transform in terms of the string collective variables $X_i$ ~\cite{Aharony:2009gg} that the action is invariant under $SO(1,D-1)$. By this symmetry the couplings Eq.\eqref{couplings1} of the four derivative term in the LW action Eq.\eqref{LWaction} are not arbitrary and are fixed in any dimension $D$ by
%--------------------------------------------- 
\begin{equation}
c_2 + c_3 = \dfrac{-1}{8\sigma}.
\label{couplings2}
\end{equation}
%---------------------------------------------

Condition \eqref{couplings2} implies that all the terms with only first derivatives in the effective string action Eq.\eqref{LWaction} coincide with the corresponding one in Polyakov-Kleinert action in the derivative expansion.
\begin{equation}
S=-\int d^2\zeta \sqrt{-{\rm{det}} g_{\alpha \beta}} \left(\sigma+\frac{1}{\alpha}(\mathcal{K}_{\alpha\beta})^2 \right), 
\end{equation}
where $g$ is the two-dimensional induced metric on the world sheet embedded in the background $\mathbb{R}^{4}$, $\alpha$ is the rigidity parameter.

%%%%%%%%%%%%%%%%%%%%%%%%%%%%%%%%%%%%%%%%%%%%%  
%%%%%%%%%%%%%%%%%%%%%%%%%%%%%%%%%%%%%%%%%%%%%  
\section{Lorentz-invariant boundary terms}
  The Lorentz symmetry which is a basic feature of YM theory dictates constraints~\cite{Aharony:2011gb,Gliozzi:2011hj,Aharony:2010cx,1126-6708-2006-05-066} over the arbitrary coefficients appearing on the effective action expansion. The string field operators of the action should be built such that these crucial symmetries are preserved. In this paper we theoretically calculate the width profile of effective open strings with boundary action made up of Lorentzian invariants. 

First we naively set out the possible form of each order in the Lagrangian density and then guided by requirement of fulfilling Lorentz symmetry the relevant terms are deduced at each coupling. We consider the case of Dirichlet boundary conditions  $X^i=0$ at both ends, pure $\zeta^0$-derivatives vanish on the boundary ($\partial_0^n X=0$). A Generic boundary action is defined as 
\begin{equation}
S_{b}=\int d\zeta^0 \left[\mathcal{L}_{1}+\mathcal{L}_2+\mathcal{L}_3+\mathcal{L}_4+...   \right],
\end{equation}
with the Lagrangian density $\mathcal{L}_i$ for each effective low-energy coupling $b_{i}$. The worldsheet coordinates have dimensions of length, a bulk term has the same order as a boundary term with one less derivative. The operator has to have an odd number $n$ of spatial derivatives $\partial^{n}_1 X$ so that it has non-vanishing value.

At the lowest order, the general allowed form is schematically proportional to $ \partial^{2} {\mathbf{X}}^{2}$ and the only possible term \cite{Luscher:2002qv} in the Lagrangian is therefore 
\begin{align}
\label{L1}
\mathcal{L}_1 = b_1 {\partial_1\mathbf{X}} \cdot {\partial_1\mathbf{X}}.
\end{align}

Consistency with the open-closed string duality~\cite{Luscher:2004ib} implies a vanishing value of the first boundary coupling $b_1=0$, as we will discuss below the Lorentz invariance requirement implies the vanishing of $b_1$ as well.

The leading order corrections due to second boundary terms with the coupling $b_2$ appears at the four derivative term in the bulk. On the boundary the general term is of the form $\partial^{3} {\mathbf{X}}^2$, apart from a term proportional to the equation of motion there is naively only one possible term  
\begin{align}
\label{L2}
\mathcal{L}_2=b_2 {\partial_0 \partial_1 \mathbf{X}} \cdot {\partial_0\partial_1 \mathbf{X}}~,
\end{align}

%%%%%%%%%%%%%%%%%%%%%%%%%%%%%%%%%%%%%%
%%%%%%%%%%%%%%%%%%%%%%%%%%%%%%%%%%%%%%
The effective action preserves the transverse rotation symmetry $SO(d-2)$, the complete $SO(1,d-1)$ Lorentz invariance is spontaneously broken by the classical solution around which we expand, but the expanded action still respects this symmetry non-linearly the rotations in $(i, j)$ plane generates
\begin{equation}
  \delta_{ij}X^{i}=\epsilon X^{j}.
  \label{rotation}
\end{equation}

In order to keep the gauge-fixing $\sigma^{1}=X^{1}$ we must make a diffeomorphism
\begin{equation}
  \begin{split}
  \delta_{12}\zeta^{1} =& \epsilon X^{2}(\zeta^{1}),\\
  \delta_{12}\zeta^{0}=&0,
  \label{reparamterization}
  \end{split}
\end{equation}

    that is, Eq.\eqref{rotation} generates rotation which should be followed by reparamterization Eq.~\eqref{reparamterization} to put the system again in the static gauge. The full nonlinear Lorentz transformation is then
\begin{equation}
  \delta^{\alpha i} X^{j}=-\epsilon\left(\delta^{ij} \zeta^{\alpha}+X^{i} \partial^{\alpha}X^{j} \right),
  \label{LT}
\end{equation}
where $\epsilon$ is an infinitesimal parameter of boosts and rotations. For each boundary action
\begin{equation}
S_{b_i}=  \int_{\partial \Sigma} d\zeta^{0} \mathcal{L}_{i}.
\end{equation}
The application of the infinitesimal transformation Eq.\eqref{LT} and requiring the action $S_{b_i}$ to vanish we obtain constraints on the values of the couplings; or realize higher-order derivatives in the choice of the action of a given coupling.

The nonlinear transformation Eq.~\eqref{LT} of the Lagrangian densities Eq.~\eqref{L2} and Eq.~\eqref{L4} generates higher-order terms at the same scaling~\cite{Aharony:2011gb,Billo:2012da} which should still acquire the same couplings. The scaling of a given term defines the order of the generated terms when applying the transformation. That is, the transform Eq.~\eqref{LT} of $\partial^{m} X^{n}$ generates terms with the same value of the difference $m-n$. The Lorentz transform of the Lagrangian density Eq.~\eqref{L2} we have two basis terms $(\partial_1 \partial_{0} X)^2$ and $\partial_1 \partial_{0} \mathbf{X} \cdot \partial_1 \mathbf{X}$
\begin{equation}
   \mathcal{L}_{2}=b_2 \sum_{k=0}^{\infty} \big[ \alpha_{k}\partial_1\partial_{0} \mathbf{X} \cdot \partial_1 \partial_{0} \mathbf{X} (\partial_1 \mathbf{X} \cdot \partial \mathbf{X})^{k}+\beta_{k+1}(\partial_{1}\partial_{0} \mathbf{X} \cdot \partial_1 \mathbf{X})^{2}(\partial_{1} \mathbf{X} \cdot \partial_1 \mathbf{X})^{k} \big].  
\end{equation}
The invariance of \eqref{L2} leads to recursion relations when solved Ref.\cite{Billo:2012da} give rise to  
\begin{equation}
     \mathcal{L}_2=b_2  \left( \dfrac{\partial_0 \partial_1 \mathbf{X} \cdot \partial_0 \partial_1 \mathbf{X}}{1+\partial_1 \mathbf{X} \cdot \partial_1 \mathbf{X}}-\dfrac{(\partial_0 \partial_1 \mathbf{X} \cdot \partial_1 \mathbf{X})^2}{(1+\partial_1 \mathbf{X} \cdot \partial_1 \mathbf{X})^2} \right).
     \label{b2Complete}
\end{equation}
However, for the third order we find then
\begin{equation}
\mathcal{L}_3= b_3 \left(\partial_1 \mathbf{X} \cdot \partial_1 \mathbf{X} \right)^2.
\label{L3}
\end{equation}
The variation of the boundary actions at first and third orders with infinitesimal nonlinear Lorentz transform Eq.\eqref{LT} of the Lagrangian densities Eq.\eqref{L1} and Eq.\eqref{L3} entails vanishing value for $b_1=0$ and $b_3=0$.

The next order, $\mathcal{L}_4$, the general effective Lagrangian on the boundary is 
\begin{equation}
\mathcal{L}_4=b_4 \partial_{0}^2\partial_{1} \mathbf{X} \cdot \partial_{0}^2\partial_{1} \mathbf{X}.
\label{L4}
\end{equation}

%The first two terms at scaling 4 is given by Ref.\cite{Billo:2012da}
The first two terms derived in Ref.\cite{Billo:2012da} are given by 
\begin{equation}
\mathcal{L}_4=b_4 \Bigg( \frac{\partial_{0}^2\partial_{1} \mathbf{X} \cdot \partial_{0}^2\partial_{1} \mathbf{X} }{1+\partial_1 \mathbf{X} \cdot \partial_1 \mathbf{X}}-\dfrac{(\partial_0^{2}\partial_1 \mathbf{X} \cdot \partial_1 \mathbf{X})^2+4 (\partial_{0}^{2}\partial_{1}\mathbf{X} \cdot \partial_{0}\partial_{1} \mathbf{X})(\partial_{0}\partial_{1} \mathbf{X})}{(1+\partial_1 \mathbf{X} \cdot \partial_1 \mathbf{X})^2}+...\Bigg).
\label{b4Complete}
\end{equation}
      
The action at the boundaries drive infinitesimal diffusion from generic source/sink into the energy density along the QCD flux tube. This causes perturbation from the free NG string behavior which is expected to evidently affect the static potential ~\cite{Billo:2011fd, Caselle:2014eka,Brandt:2017yzw,Bakry:2017fii,Bakry:2020ebo,Bakry:2020flt} and the profile near the color sources, short separation distances as well as high temperatures~\cite{Bakry:2017fii,Bakry:2020ebo,Bakry:2020flt, Bakry:2018kpn} and excited spectrum~\cite{Bicudo:2017uyy}. 
  In the next section, we shall lay out the perturbative expansion of the boundary action and estimate the subsequent augmentation/lessening of mean-square width of the effective string in $D$ dimension at any temperature.
   
%%%%%%%%%%%%%%%%%%%%%%%%%%%%%%%%%%%%%%
%%%%%%%%%%%%%%%%%%%%%%%%%%%%%%%%%%%%%%
\section{The boundary terms contribution to the  energy width}
\label{Section2}   

Let us consider the free NG action around which the perturbative expansion is taken
\begin{equation}
  S_{\ell o}[\mathbf{X}]=S_{cl}+\dfrac{\sigma}{2} \int d\zeta^{1} \int d\zeta^{0}  \Big(\dfrac{\partial \mathbf{X}}{\partial \zeta^{\alpha}} \cdot \dfrac{\partial \mathbf{X}}{\partial \zeta^{\alpha}}\Big),
  \label{LOaction}
\end{equation}
%---------------------------------------------
%---------------------------------------------
The next-to-leading NG term combined with the expansion of the surface term on the boundary
\begin{equation}
\begin{split}  
\label{Pert}  
S^{\rm{Pert}}[\mathbf{X}]&=S_{n\ell o}+S_b,\\
&= \left(S_{n \ell o}+S_{b_{2}}+S_{b_{4}}+..\right)+.... , 
\end{split}
\end{equation}
defines the perturbation from the effective LW action Eq.~\eqref{LWaction} with higher-order geometrical terms left out~\footnote{The extrinsic curvature correction to the width profile has been worked out elsewhere in Ref.~\cite{Bakry:2017fii,Bakry:2020ebo,Bakry:2020flt}}. 

The mean-square width of the string is defined as the second moment of the field with respect to the center of mass of the string $X_0$ and is given by    
\begin{equation}
W^2(\zeta)=\dfrac{\int  DX (\mathbf{X}(\zeta^1,\zeta^0)- \mathbf{X_0} )^2 e^{-S[\mathbf{X}]}} {\int D\mathbf{X} e^{-S[\mathbf{X}]}}, 
\label{Width2}
\end{equation}

Expanding around the free-string action Eq.~\eqref{LOaction} the squared width of the string ~\cite{Gliozzi:2010zt} is given by
\begin{equation}
\begin{split}
W^2(\zeta^{1}) &= W_{\ell o}^2(\zeta^{1})-\langle \mathbf{X}^{2}(\zeta^{0},\zeta^{1}) S^{\rm{Pert}} \rangle_0 +2 \gamma \langle\partial_{\alpha} \mathbf{X}^{2}(\zeta^{0},\zeta^{1})\rangle_0 +  \gamma^2 \langle \partial_{\alpha}^{2} \mathbf{X}^{2}(\zeta^{0},\zeta^{1})\rangle_0 \\ 
&-\dfrac{\gamma^2}{L_{T} R} \int d\zeta^{0}~d\zeta^{1}~d\zeta^{0'}~d\zeta^{1'}  \langle  \partial_{\alpha}^{2} \mathbf{X}(\zeta^{0},\zeta^{1}) \cdot \partial_{\alpha^\prime}^{2}  \mathbf{X}(\zeta^{0'},\zeta^{1'})\rangle_0. 
\end{split}
\label{TwoLoopExpansion}
%\end{align}
\end{equation}
where the vacuum expectation value $\langle ...\rangle_0$  is with respect to the free-string partition function, $\gamma$ is an effective low energy parameter and $L_{T}$ is the length of compactified time direction for cylindrical boundary condition. 

Substituting Eq.~\eqref{Pert} in the low-energy parameter expansion Eq.~\eqref{TwoLoopExpansion} the mean-square width 
%---------------------------------------------  
\begin{equation}
\begin{split}
W^2 (R,L_{T})=&\left\langle(\mathbf{X}^2(\zeta^1,\zeta^0) S_{\ell o})\right\rangle-\left\langle(\mathbf{X}^{2}(\zeta^{1},\zeta^{0}) S_{n\ell o}+\mathbf{X}^{2}(\zeta^{1},\zeta^{0})S_{b_2}+\mathbf{X}^{2}(\zeta^1,\zeta^{0})S_{b_4}+...) \right\rangle,\\\nonumber
=& W^2_{\ell o}(R,L_{T})+ W_{n\ell o}^{2}+W^2_{b}(R,L_{T})+...\\
\end{split}  
\end{equation}
with the contribution of the boundary action
\begin{equation}
W^2_{b}(R,L_{T})=W^2_{b_2}(R,L_{T}) +W^2_{b_4}(R,L_{T}),\nonumber
\label{eq:W2_Pert}
\end{equation}
and the mean square width of the free string
\begin{equation}
W^2_{\ell o}(R,L_{T})=\dfrac{\int  DX ( \mathbf{X}(\zeta^{1},\zeta^0))^2 e^{-S_{\ell o}[\mathbf{X}]}} {\int DX e^{-S_{\ell o}[\mathbf{X}]}}. 
\end{equation}
The Green-function defines the two-point free propagator
\begin{equation}
G\left(\zeta^0,\zeta^1; \zeta^{0'},\zeta^{1'}\right)= \langle \mathbf{X}(\zeta^{0},\zeta^1 ) \cdot \mathbf{X}(\zeta^{0'},\zeta^{1'})\rangle,
\end{equation}
which is the solution of Laplace equation on a cylindrical sheet~\cite{Gliozzi:2010zt} of surface area $RL_{T}$  and is given in spectral form ~\cite{Gliozzi:2010zt} by
%---------------------------------------------  
\begin{equation}
G(\mathbf{\zeta;\zeta'})=  \frac{1}{\pi  \sigma }\sum _{n=1}^{\infty } \frac{1}{n \left(1-q^n\right)}\sin \left(\frac{\pi  n \zeta^{1}}{R}\right) \sin \left(\frac{\pi  n \zeta^{1'}}{R}\right)  \left(q^n e^{\frac{\pi  n (\zeta^0-\zeta^{0'})}{R}}+e^{-\frac{\pi n (\zeta^0-\zeta^{0'})}{R}}\right),
\label{GreenPropagator}
\end{equation}
%---------------------------------------------
  where $q=e^{\frac{-\pi L_{T}}{R} }$ is the complementary nome. The Dirichlet boundary condition corresponding to fixed displacement vector at the ends $\zeta^{1}=0$ and $\zeta^{1}=R$ and periodic boundary condition in time $\zeta^{0}$ with period $L_{T}$ ~\cite{Gliozzi:2010zt} are encoded in the above propagator. 

The expectation value ~\cite{Caselle:1995fh,allais,Gliozzi:2010zv} of the mean-square width corresponds to the Green-function correlator of the free bosonic string theory in two dimensions
% In $D$ dimension and for cylindrical boundary conditions Eq.~\eqref{bc1} and Eq.~\eqref{bc2} the mean-square width would read
\begin{equation}
W^{2}_{{\ell o}}(\zeta_1,\tau) = \frac{D-2}{2\pi\sigma_{0}}\log\left(\frac{R}{R_{0}(\zeta_1)}\right)+\frac{D-2}{2\pi\sigma_{0}}\log\left| \,\dfrac{\vartheta_{2}(\pi\,\zeta_1/R;\tau)} {\vartheta_{1}^{\prime}(0;\tau)} \right|,
\label{W2LO}
\end{equation}
%--------------------------------------------- 
where $\theta$ are Jacobi elliptic functions
%--------------------------------------------- 
\begin{eqnarray}
\theta_{1}(\zeta;\tau)=2 \sum_{n=0}^{\infty}&(-1)^{n}q_1^{n(n+1)+\frac{1}{4}}\sin((2n+1)\,\zeta),\nonumber\\
\theta_{2}(\zeta;\tau)=2 \sum_{n=0}^{\infty}&q_1^{n(n+1)+\frac{1}{4}}\cos((2n+1)\zeta),
\end{eqnarray}
%---------------------------------------------
\noindent with  $q_1=e^{\frac{i \pi}{2}\tau}$, and  $R^{2}_{0}$ is the UV cutoff which has been generalized to be dependent on distances from the sources. The above formula Eq.~\eqref{W2LO} can be pinned down through the standard relations between the elliptic Jacobi and Dedkind $\eta$ functions to either of the equivalent forms derived in Refs.~\cite{Gliozzi:2010zt,Bakry:2017fii,Bakry:2020ebo,Bakry:2020flt}.  

  The NLO term from the low energy parameter expansion~Eq.\eqref{eq:W2_Pert} to the width of NG string have been worked out in detail  \label{eq:W2_Pert} in Ref.\cite{Gliozzi:2010zt}, the width due to the self-interaction is modified by
\begin{equation}
\begin{split}
  W^{2}_{n\ell o}=&\dfrac{(D-2)\pi}{12\sigma^2 R^2}\Big\{\tau \Big(q \frac{d}{dq} -\frac{D-2}{12}E_2(\tau)\Big)\left[E_2(2\tau)-E_2(\tau)\right]-\frac{D-2}{8 \pi} E_2(\tau)\Big\}\\
&+\frac{\pi}{12 \sigma R^2}\left[E_2(\tau)-4E_2(2\tau)\right]\left(W_{lo}^2-\frac{D-2}{4\pi \sigma}\right).
\end{split}
\label{SI}
\end{equation}

In the following we evaluate the correction to the mean-square width of the free string with Lorentz invariant boundary action at these two orders. We consider the expectation value of the quadratic field excitation up to the leading order term in the boundary action~\eqref{Sb2}.
%-------------------------------------------------

The perturbative expansion of the boundary action $S_{b_{2}}$ corresponding to the Lagrangian density Eq.\eqref{L2} is given by
\begin{equation}
S_{b_{2}}= b_2  \int_{\partial \Sigma} d\zeta^0  \left(\partial_0 \partial_1 \mathbf{X} \cdot \partial_0 \partial_1 \mathbf{X}-\left(\partial_0 \partial_1 \mathbf{X} \cdot \partial_0 \partial_1 \mathbf{X}\right) \left(\partial_1 \mathbf{X} \cdot \partial_1 \mathbf{X} \right)+...\right).
\label{Sb2}
\end{equation}
The generic form of the Wick-contracted term representing the perturbative contribution of the boundary term to the expectation value is 
  %-----------------------------------------------
\begin{equation}
\langle X^2 S_{b2} \rangle =b_2 (\langle (\mathbf{X} \cdot \mathbf{X}) \partial_{0}^{2} \partial_{1}^{2} (\mathbf{X} \cdot \mathbf{X})\rangle+\langle \partial_{0}\partial_{1}(\mathbf{X} \cdot \mathbf{X}) \partial_{0} \partial_{1} (\mathbf{X} \cdot \mathbf{X})\rangle).
\label{Expectation1}
\end{equation}
Upon point-splitting the expectation value in terms of Green propagators the mean-square width becomes
\begin{equation}
  W^2_{b_2}=-(D-2)b_{2}  \int_{\partial \Sigma} d \zeta^{0'} \lim_{\substack{\zeta' \to \zeta}} \big[ \partial_{0} \partial_{1'} G(\mathbf{\zeta;\zeta'})\partial_{0'} \partial_{1} G(\mathbf{\zeta';\zeta})+ G(\mathbf{\zeta;\zeta'}) \partial_{0} \partial_{1} \partial_{0'} \partial_{1'} G(\mathbf{\zeta';\zeta}) \big].
  \label{W2b2}
\end{equation}
%---------------------------------------------
the above two terms define the sum of the expectation values
\begin{equation}
W^{2}_{b_2}=-\left(W^{2}_{1,b_2}+W^{2}_{2,b_2}\right).  
\label{W1b2W2b2}
\end{equation}
Substituting the free Green propagator Eq.\eqref{GreenPropagator} in the above expectation value. We perform the integrals over $\zeta^{0}$ after evaluating the derivatives and rewriting the resulting summations in closed form.

%------------------------------
The evaluation of the two correlators Eq.\eqref{W1b2W2b2} involve cumbersome manipulations, we present the detailed calculus in Appendix(B). The $\zeta$-function regularization of the divergent sums is performed on each expectation value appearing in the limits, $\epsilon, \epsilon' \to 0 $. The mean-square width  $W^{2}_{b_2}$ turn out to be

 \begin{equation}
   \begin{split}
   W^{2}_{b_2}(R,L_{T})=& \frac{-\pi^{3} b_2 (D-2)}{16 R^5 \sigma^2}\Bigg(\frac{1}{3}E_{2}(2i\tau) +2\frac{\vartheta _1^{\prime }}{\vartheta _1}\left(\frac{i\pi }{2}\tau,q\right)^{(1)}-\frac{91}{6}\Bigg).
   \end{split}
   \label{1wb2}
\end{equation}

   In  Wilson loops~\cite{Caselle:2010pf, Billo:2011fd, Billo:2012da} operators the boundary action survives over both the spatial and temporal extends, the boundary action is given by     
\begin{equation}
  S_{b_{2}}= b_2  \int_{\partial \Sigma_t} d\zeta^0  \left(\partial_0 \partial_1 \mathbf{X} \cdot \partial_0 \partial_1 \mathbf{X}\right) + b_2 \int_{\partial \Sigma_s} d\zeta^1 \left(\partial_0 \partial_1 \mathbf{X} \cdot \partial_0 \partial_1 \mathbf{X}\right),  
\label{S2b2Wilson}
\end{equation}
  this surface term lives on the boundary $\partial \Sigma$ which in the action Eq.~\eqref{S2b2Wilson} is conveniently chosen as a rectangle-shaped Wilson's loop circumfering the spatial-temporal area of $R \times L$, the curves $\partial\Sigma_t$ and $\partial\Sigma_s$ stands for temporal and spatial parts of the loop, respectively.

  The direct calculation of the expectation value of the $\langle S_{b} \rangle$ entails contributions from both the temporal and spatial parts to the static quark potential~\cite{Billo:2011fd}. As a consequence of the symmetry of the propagator Eq.~\eqref{GreenPropagator}, these two corrections give similar formulas~\cite{Billo:2011fd}; however, with the role  between the source separation, $R$, and temporal extent, $L$, exchanged. Moreover, the contribution from the temporal path is formally equivalent~\cite{Billo:2011fd} to that of two Polyakov loop with the identification of the temporal height of the Wilson's loop $L$ instead of the cylindrical time extend $L_{T}\to L$. 

  The generalization of the expectation value of the width due to two Polyakov loops Eq.\eqref{2wb2} to that due to the Wilson's loop is accordingly evaluated as the expectation value of
\begin{equation}
 \mathcal{W}^{2}_{b_2}=\langle X^{2} S_{b_2} \rangle_{\partial \Sigma_t} + \langle X^{2} S_{b_2} \rangle_{\partial \Sigma_s},
\end{equation}
for the boundary action given by Eq.\eqref{S2b2Wilson}.  The spatial part of this expectation value is an integral of the diffused energy of the string along the spatial path the infinitesimal-time interval of either the creation/annihilation of the static quark-antiquark pair. As  the mean-square width can be deduced to have the form    
\begin{equation}
%\begin{split}
  \mathcal{W}^{2}_{\square,b_2}(\tau;b_{2})=W^{2}_{b_2}\left(R,L_{T}\right)+W^{2}_{b_2}\left(L_{T},R\right)
%  \frac{-\pi b_2 (D-2)}{4 \sigma^2} \left[\frac{1}{R^3} \left(\frac{1}{8}-\frac{1}{24}E_{2}\left(\frac{i L}{R}\right)\right)+ \frac{1}{L^3} \left(\frac{1}{8}-\frac{1}{24} E_{2}\left(\frac{i R}{L}\right) \right) \right],
%\end{split}
\label{2wb2}
\end{equation}
where $L$ is the temporal extend of Wilson loop. 

 The next Lorentz invariant non-vanishing Lagrangian density is Eq.\eqref{L4} corresponding to the coupling $b_4$, we consider the leading term in the expansion of boundary action  
\begin{equation}
S_{b_{4}}= b_4 \int_{\partial \Sigma} d\zeta^0  \left(\partial_0 \partial_1 \mathbf{X} . \partial_0 \partial_1 \mathbf{X}-\left(\partial_0 \partial_1 \mathbf{X} . \partial_0 \partial_1 \mathbf{X}\right) \left(\partial_1 \mathbf{X} .\partial_1 \mathbf{X} \right)+...\right).
\label{Sb2}
\end{equation}
  The Wick-contraction of the operator $X^2$ with the leading term yields the expectation value 
\begin{equation}
\langle \mathbf{X}^2 S_{b4} \rangle = b_4 \langle (\mathbf{X} \cdot \mathbf{X}) \partial_{0}^{2} \partial_{1}^{2} (\mathbf{X}\cdot \mathbf{X})\rangle+\langle \partial_{0}\partial_{1}(\mathbf{X} \cdot \mathbf{X}) \partial_{0} \partial_{1} (\mathbf{X} \cdot \mathbf{X})\rangle.
\label{Expectation1}
\end{equation}

 The expectation value Eq.\eqref{Expectation1} in terms of Green propagators after point-splitting become 
\begin{equation}
  W^2_{b_4}=-(D-2)b_{4}  \int_{\partial \Sigma}  d\zeta^{0'} \lim_{\substack{\zeta' \to \zeta}} \left( \partial_{0} \partial_{1'} G(\mathbf{\zeta;\zeta'})\partial_{0'} \partial_{1} G(\mathbf{\zeta';\zeta})+ G(\mathbf{\zeta;\zeta'}) \partial_{0} \partial_{1} \partial_{0'} \partial_{1'} G(\mathbf{\zeta';\zeta}) \right).
  \label{eq:Integral}
\end{equation}

   Similarly, substituting the free propagator Eq.\eqref{GreenPropagator} in the above expectation value and integrating over $\zeta^{0}$, see Appendix(C) for details. The expectation value of the fourth-order boundary term in the action 
\begin{equation}
  W^{2}_{b_4}=-\left(W^{2}_{1,b_4}+W^{2}_{2,b_4}\right),
\label{W2b4}  
\end{equation}
 which turn up as,
\begin{equation}
%\begin{split}
  W^{2}_{b_4}(R,L_{T})
 % =&\frac{\pi^3 (D-2)b_4 }{32 R^5 \sigma^2} \Bigg[\Bigg(\left( \frac{32}{240}\left(E_{4}(2\tau)-1\right)-\frac{16}{120} \right)-\left( \frac{8}{24}\left(1-E_{2}(2\tau)\right)+\frac{1}{12}\right)+h\left(\frac{\tau}{2} \right)\Bigg)\Bigg]\\
  =\frac{\pi^5 (D-2)b_4 }{32 R^7 \sigma^2} \Bigg(  \frac{2}{15}E_{4}\left(2i\tau \right)+\frac{1}{3}E_{2}\left(2i\tau \right)-\frac{7}{60}-2\frac{\vartheta _1^{\prime }}{\vartheta _1}\left(\frac{i\pi }{2}\tau,q\right)^{(1)}-32\Bigg),
  %\end{split}
\label{wb4}  
\end{equation}

\begin{figure*}[!hpt]	
	\begin{center}						
		\subfigure[]{\includegraphics[scale=0.77]{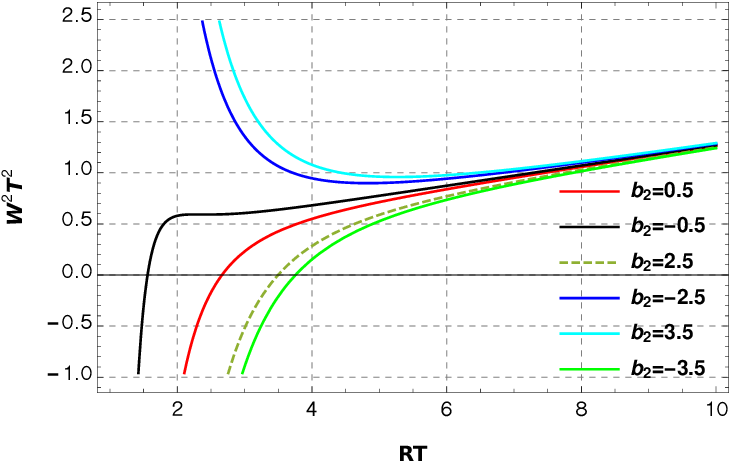}}
		\subfigure[]{\includegraphics[scale=0.77]{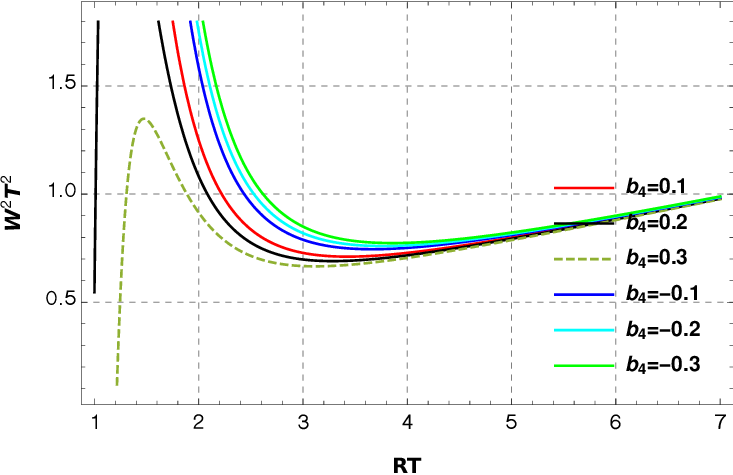}}
		%\subfigure[]{\includegraphics[scale=0.37]{stringWidth_WilsonBounds_NGlo_BnloLta8}}
		%\subfigure[]{\includegraphics[scale=0.37]{stringWidth_WilsonBounds_NGnlo_BnloLta8}}
                \caption{The plot compares the mean-square width of the effective string bounded by two Polyakov-loops at the depicted values of the boundary coupling $b_2$ and  $b_4$ in the middle plane.The contribution of the boundary action to the width profile $W^2(R/2)=W^2_{\rm{NG_{(\ell o)}}}(R/2)+W^2_{b}$ of the leading order NG string; however, at temperature scale $L_{T}a^{-1}=6$.}
		\label{LOP}		
	\end{center}
\end{figure*}

\begin{figure*}[!hpt]	
	\begin{center}						
		\subfigure[]{\includegraphics[scale=0.75]{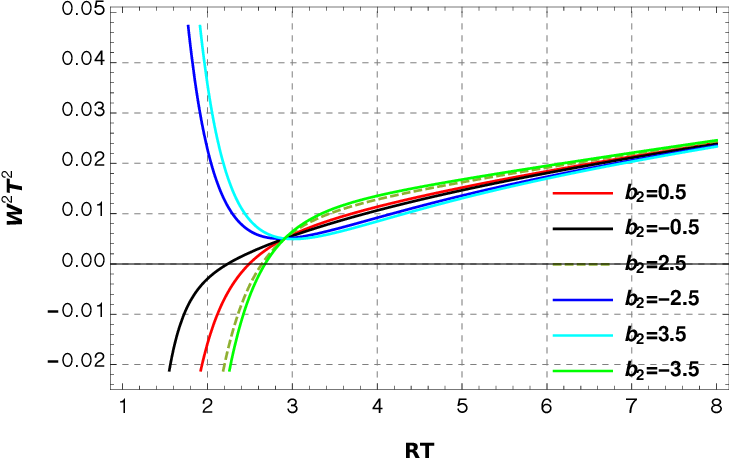}}
		\subfigure[]{\includegraphics[scale=0.75]{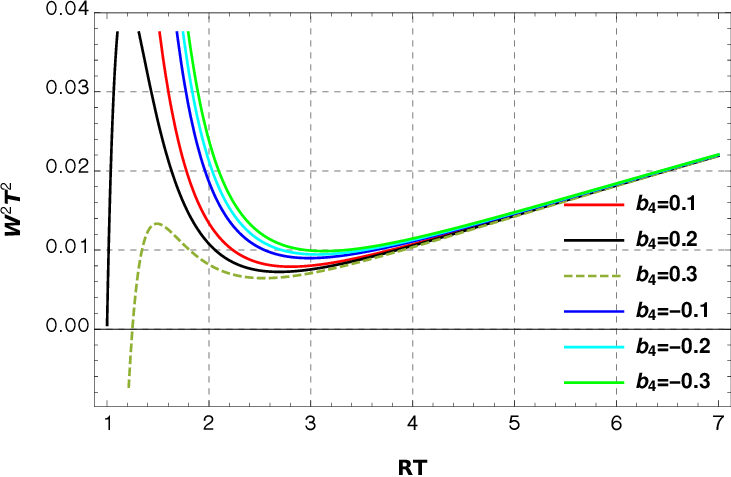}}
		%\subfigure[]{\includegraphics[scale=0.37]{stringWidth_WilsonBounds_NGlo_BnloLta8}}
		%\subfigure[]{\includegraphics[scale=0.37]{stringWidth_WilsonBounds_NGnlo_BnloLta8}}		
		\caption{The plot compares the mean-square width of the effective string bounded by two Polyakov-loops at the depicted values of the boundary coupling $b_2$ and  $b_4$ in the middle plane.The contribution of the boundary action to the width profile $W^2(R/2)=W^2_{\rm{NG_{(n\ell o)}}}(R/2)+W^2_{b}$ of the next to leading order NG string at zero temperature.}
		\label{LOandNLO}		
	\end{center}
\end{figure*}

On the otherhand, the corresponding boundary corrections of the width of Wilson loop  at coupling $b_4$ can be deduced following up the same line of reasoning leading to Eq.\eqref{2wb2}. The  next-to-leading boundary correction for Wilson loop of rectangular area $L\times R$ is, accordingly, given by
\begin{equation}
\begin{split}
  \mathcal{W}^{2}_{\square;b_4}(\tau;b_4)=W^{2}_{b_4}(R,L_{T})+W^{2}_{b_4}\left(L_{T};R\right)
%  \frac{\pi^3 (D-2)b_4}{32 \sigma^2} \Bigg[\frac{1}{R^5}\left(\,\text{E}_2\left(\frac{iL}{2 R}\right)-\frac{5}{4}\right) \left(\frac{11 }{36}\text{E}_2\left(\frac{iL}{4 R}\right)-\frac{5 }{9}\text{E}_2\left(\frac{iL}{2 R}\right)-\frac{55}{12}\right)\\
%  &+\frac{1}{L^5} \left(\,\text{E}_2\left(\frac{iR}{2 L}\right)-\frac{5}{4}\right) \left(\frac{11 }{36}\text{E}_2\left(\frac{iR}{4 L}\right)-\frac{5 }{9}\text{E}_2\left(\frac{iR}{2 L}\right)-\frac{55}{12}\right) \Bigg].\\
  \end{split}
\label{2wb4}
\end{equation}

  The leading non-vanishing boundary corrections to the flux-tube width Eq.\eqref{1wb2} indicate an inverse decrease with the third power of the length scale $R$. This suggests  effects that are more noticeable near the intermediate and short string length scale. In Fig.\ref{LOandNLO} the mean-square width $W^{2}$, of the free NG string Eq.\eqref{W2LO} and self-interacting NG string Eq.\eqref{SI} togather with the boundary corrections Eq.\eqref{1wb2},  are plotted at the middle plane versus the string length such that

\begin{equation}
  W^2(R/2)=W^2_{\rm{\ell o}}(R/2)+W^2_{n\ell o}+W^2_{b_2}+W_{b_4}^{2},
  \label{all}
\end{equation}
 
  On Fig.\ref{LOP} we compare modifications in the width profile which were received from the depicted positive/negative values of the coupling parameter $b_2$ and $b_4$ in Eq.\eqref{1wb2} and Eq.\eqref{wb4}.

The boundary term $S_{b_2}$ in the effective string action may increase or decrease the mean-square width of NG string depending on whether positive or negative values of the coupling parameter $b_2$ are considered. However, the resultant effects on the width is a compromise between the value of the two couplings $b_2$ and $b_4$ and the corresponding signs.

As expected from the dimensional considerations the remarkable effects of the corrections occur over the intermediate and short string length scale. The diffusion of the interaction from the sources at the boundaries along the string is more stringent for relatively short string length. For the values of $b_4$ considered in Fig.\ref{LOP} the $S_{b_{4}}$ boundary term in the action appears to dominate and can in principle fine tune the width at shorter distances.

 \begin{figure*}[!htb]	
	\begin{center}						
		\subfigure[]{\includegraphics[scale=0.85]{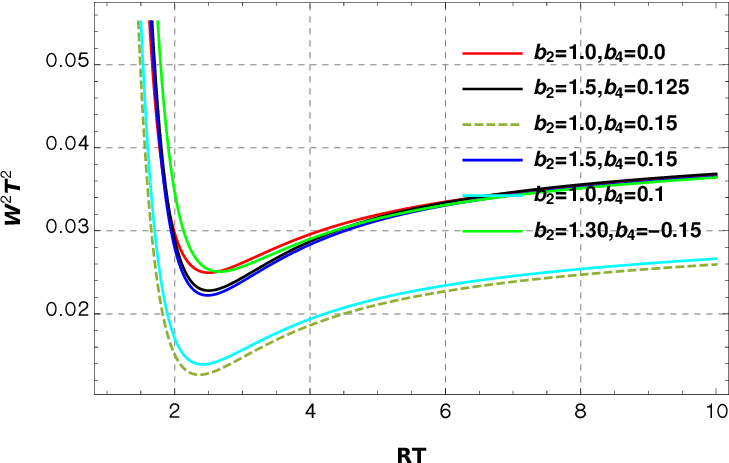}}
%		\subfigure[]{\includegraphics[scale=0.5]{Fig2b.eps}}
		%\subfigure[]{\includegraphics[scale=0.37]{stringWidth_WilsonLoopRxL}}
		%\subfigure[]{\includegraphics[scale=0.37]{stringWidth_PolyakovLoopsRxL}}
		\caption{Compares the profile of the effective string at next to leading order in NG action for the depicted values of the couplings $b_2$ and $b_4$  ($W^2=W^2_{\rm{n\ell o}}(R/2)+\mathcal{W}^2_{b_2}+\mathcal{W}_{b_4}^{2}$) of Wilson loop of area $R \times L$ with $L {a}^{-1}=4$ time slices at zero temperature $T=0$.}
		\label{b4}		
	\end{center}
\end{figure*}

The comparison of the mean-square width of the free NG string with the corresponding self-interacting string for a fixed value of the coupling $b_2$ and $b_4$ reveals subtle dilution of the boundary action effects if considered with the associated self-interaction of NG string. This suggests diffusion attenuated to some extend with the viscose medium driven by the self-interacting string field.

In Fig.\ref{b4}(a) we plot the width of the effective string versus the source separation and zero temperature in the middle plane of the string. The plot compares the width profile of the self-interacting string -Eq.\eqref{W2LO} and Eq.\eqref{SI}- setup by Wilson loop
\begin{equation}
W^2(R/2)=W^2_{\rm{n\ell o}}(R/2)+\mathcal{W}^2_{\square,b_2}+\mathcal{W}_{\square,b_4}^{2},
\label{all}
\end{equation}
  with $\mathcal{W}^{2}_{\square,b_2}$ and $\mathcal{W}^{2}_{\square,b_4}$ are given by Eq.\eqref{2wb2} and Eq.\eqref{2wb4}, respectively. The first term in Eq.\eqref{all} signifies the zero temperature logarithmic prodening of the effective string. A similar plot in Fig.\ref{b4}(b) illustrates; however, the perturbative mean-square width with the decrease of the temperature Eq.\eqref{all}. The plot shows the convergence of the solutions in the limit of infinite temporal $L_{T} \to \infty$ extend. 

The corrections provided by the boundary action to static $Q\bar{Q}$ potential seem to explain to some extend the deviations appearing when constructing the static mesonic states with Polyakov loop correlators~\cite{Brandt:2010bw}. The  $Q\bar{Q}$ potential is well described by the up to surprisingly small distances as $R=0.3$ fm using the boundary action at temperature near the end of QCD Plateau $T/T_c=0.8$ fm. At a higher temperature, the inclusion of the boundary corrections up to the fourth order $b_4$ together with string rigidity has been found ~\cite{Bakry:2018kpn} to be viable in providing good fits for distances as small as $R=0.5$ fm.

The lattice simulations of the energy-density profile is a more challenging quantity to measure compared to the $Q\bar{Q}$ potential. The exponentially decaying field corelators involve the field strength tensor leading ~\cite{PhysRevD.82.094503, Bakry:2010sp, Bicudo:2017uyy, deForcrand:2005vv} to a substantial numerical effort ~\cite{Gliozzi:2010zv,Gliozzi:2010zt,Bakry:2017fii,Bakry:2020ebo,Bakry:2020flt,Bakry:2018kpn} to attain a good enough precision.
   
   The boundary corrections to the mean-square width of the flux-tube dictated by Eq.\eqref{W2b2} could be relevant to fine deviations such as those famed to occur on intermediate source separation scale. The detection of these subtleties is, of course, subject to the continuous improvement in the resolution of the lattice data and computational power. Nevertheless, one would expect an impact of these terms if to carry out the numerical investigations with the Wilson loop operators~\cite{Caselle:1995fh}.
   
%%%%%%%%%%%%%%%%%%%%%%%%%%%%%%%%%%%%%
%%%%%%%%%%%%%%%%%%%%%%%%%%%%%%%%%%%%%   
\section{Summary and Conclusion}
In this work, the width of the energy profile of string bounded by two Polyakov loops is derived as a function of the temperature in any dimension $D$. We have considered LW string with two Lorentz invariant subleading boundary terms in the effective action. 

The mean-square width have been estimated as a low energy perturbative expansion around the free string action. The width is derived for open strings with Dirichlet boundary condition on cylinder. We have implemented the technique of the $\zeta$-function regularization to the quadratic operators in the corresponding expectation values of the mean-square width of the string.

%%%%%  The Green functions have been evaluated with open-string propagator on cylinder, i.e, Dirichlet boundary condition. 

The main findings in the present paper is the boundary correction to mean-square width of the flux-tube constructed by two Polyakov loops which given by Eq.\eqref{1wb2} at the coupling order $b_2$ of the Lagrangian density at the boundaries. The corrections induced by the next Lorentz invariant term at coupling $b_4$ have been estimated as well and are given in Eq.\eqref{W2b4}. The expectation values of two Polyakov loops can be directly generalized to Wilson's loop, where the boundary action survives in both the spatial and temporal extents, are given by Eq.\eqref{2wb2} and Eq.\eqref{2wb4} at coupling  $b_2$ and $b_4$, respectively.

  The fine structure of the mean-square width dictated by Eq.\eqref{1wb2} and Eq.\eqref{W2b4} indicates effects at the fifth order in the inverse length scale $1/R^5$. This could be relevant to QCD strings at relatively small quark separations but where the long effective string theory still holds.
%  as well as the hot phase of confining gauge groups. 

  It would be interesting to investigate the width profile taking into account the boundary action in the numerical simulation of Abelian and non-Abelian ~\cite{Khalil:2022idw,Khalil:2022gpj,Bakry:2011cn,Bakry:2010zt,Bakry:2015csa,bakry:178,Caselle:2014eka,Caselle:2004jq,Caselle:2016mqu} gauge groups. In particular, the generalization of the calculation presented in this paper to Wilson loops~\cite{Khalil:2022gpj,Caselle:2010pf,Billo:2011fd,Billo:2012da,Bonati:2018uwh} operators where more impact of the boundaries is expected. The excited meson~\cite{Bakry:2020xcu,Bicudo:2018yhk,Majumdar:2002mr} or baryonic configurations~\cite{Bakry:2020xcu,Alexandrou:2002sn, Alexandrou:2001ip, Borisenko:2018zzd, Bakry:2016aod, Bakry:2014gea, Bissey, Koma:2017hcm} are very relevant system to discuss the boundary corrections. 

  %%%%%%%%%%%%%%%%%%%%%%%%%%%%%%%%%%%%%%

\acknowledgments
    This work has been supported by the Chinese Academy of Sciences President's International Fellowship Initiative grants No.2015PM062 and No.2016PM043, the Recruitment Program of Foreign Experts, the Polish National Science Centre (NCN) grant 2016/23/B/ST2/00692, NSFC grants (Nos.~11035006,~11175215,~11175220), the Hundred Talent Program of the Chinese Academy of Sciences (Y101020BR0). 
%%%%%%%%%%%%%%%%%%%%%%%%%%%%%%%%%%%%%%
\appendix
\section{Appendix: Functions and Identities}

\underline{Zeta Regularization}\\
	In the following,  we quote the partition some of basic properties of the zeta-regularized products~\cite{Yoshimoto2002}.
 	Let us consider the convergence index which is the least integer $h$ such that the series for $Z[h+ 1]$ converges absolutely.\\
        \begin{itemize}
	\item Partition Property.\\	
 	Let $\{\varphi^1 \}_k$ and $\{\varphi ^2\}_k$ 
 	be $\bm{\zeta}$- 
 	and let regularizable sequences $\{\varphi \}_k=\{\varphi^1\}_k \cup  \{\varphi ^2\}_k$ disjoint union then
 	\begin{equation}
 	\prod_{k} \varphi_k =\prod_{k} \{ \varphi^1 \}_k \prod_{k} \{\varphi ^2\}_k
 	\end{equation}
 	\item Splitting Property.\\
 	If $\{\varphi _k\}$ is zeta-regularizable and $\prod_k a_k $ is convergent absolutely then
 	\begin{equation}
 	\prod_{k} a_k \varphi_k= \prod_k a_k \prod_{k} \varphi_k
 	\end{equation}
		
\end{itemize}  
\paragraph{Identity I}
The Jacobi elliptic $\vartheta_1$ function~\cite{whittaker_watson_1996} of the first kind is defined as 
\begin{eqnarray}
\vartheta_1(z,q)&=& 2 \sqrt[4]{q} \sin (z) \prod _{k=1}^{\infty } \left(1-q^{2 k}\right) \left(1-2 q^{2 k} \cos (2 z)+q^{4 k}\right),\label{theta1}\\
\vartheta_2(z,q)&=& 2 \sqrt[4]{q} \cos (z) \prod _{k=1}^{\infty } \left(1-q^{2 k}\right) \left(1+2 q^{2 k} \cos (2 z)+q^{4 k}\right),\label{theta2}\\
\vartheta_3(z,q)&=& \prod _{k=1}^{\infty } \left(1-q^{2 k}\right)                       \left(1+2 q^{2 k-1} \cos (2 z)+q^{4 k-2}\right),\label{theta3}\\
\vartheta_4(z,q)&=& \prod_{k=1}^{\infty } \left(1-q^{2 k}\right)                        \left(1-2 q^{2 k-1} \cos (2 z)+q^{4 k-2}\right),\label{theta4}\\
\label{theta}
\end{eqnarray}
  with the nome defined as $q=e^{i \pi  \tau }\land \Im(\tau )>0$.

\paragraph{Identity II} The logarithmic derivatives of $\vartheta_1(z,q)$ can be expressed as a series sum of hyperbolic functions as
\begin{equation}
\sum _{k=-\infty}^{\infty } \coth(\pi  k \lambda +z)=\frac{\vartheta _1^{\prime}\left(i z,e^{-\pi  \lambda }\right)}{\vartheta _1\left(i z,e^{-\pi  \lambda }\right)}.
\label{IdentityII}
\end{equation}
 
\paragraph{Identity III} The series expansion of the derivative of $\frac{\vartheta_1^{\prime}}{\vartheta_{1}}(z,q)$

\begin{equation}
\sum _{k=-\infty}^{\infty } \text{csch}(\pi  k \lambda +z)^2=\frac{i\vartheta _1^{\prime \prime}\left(i z,e^{-\pi  \lambda }\right)}{\vartheta _1\left(i z,e^{-\pi  \lambda }\right)}- \frac{\vartheta _1^{\prime }\left(i z,e^{-\pi  \lambda }\right)^2}{\vartheta _1\left(i z,e^{-\pi  \lambda }\right)^2}.
\label{IdentityIII}
\end{equation}

%\paragraph{Identity III} The square of logarithmic derivative~\cite{Zhao}

%\begin{equation}
%\left(\frac{\vartheta _1^{\prime }(z,q)}{\vartheta _1(z,q)}\right)^2=\frac{\vartheta _1^{\prime \prime}(z,q)}{\vartheta _1(z,q)}-\left(\frac{\vartheta _1^{\prime }(z,q)}{\vartheta _1(z,q)}\right)^{\prime}.
%\label{IdentityIII}
%\end{equation}

\paragraph{Identity IV}
  The modular $\pi \tau$ increase to the argument of $\vartheta _1$ function has the property~\cite{whittaker_watson_1996}
\begin{equation}
\vartheta _1(\pi  \tau +z,q)=-\frac{e^{-2 i z} }{q} \vartheta _1(z,q).
\label{IdentityIV:eq1}
\end{equation}

For $m\in \mathbb{Z}$
\begin{equation}
\vartheta _1(z+m \pi  \tau ,q)=(-1)^m q^{-m} \vartheta _1(z,q) \exp (-i (\pi  (m-1) m \tau +2 m z)).\\
\label{IdentityIV:eq2}
\end{equation}

  The modular increase of the logarithmic derivative can be deduced from properties of $\vartheta_1$ function, for $m=1$ we obtain
\begin{equation}
\frac{\vartheta _1^{\prime }(z\pm{\pi}  \tau ,q)}{\vartheta _1(z \pm{\pi}  \tau ,q)}=\frac{\vartheta _1^{\prime }(z,q)}{\vartheta _1(z,q)}\pm{2} i,
\label{IdentityIV:eq3}
\end{equation}
and $m=-1,+1$
\begin{equation}
  \frac{\vartheta _1^{\prime \prime}(z\pm{\pi}  \tau ,q)}{\vartheta _1(z \pm{\pi}  \tau ,q)}=\frac{\vartheta^{\prime\prime} \left(z,q \right)}{\vartheta _1\left(z,q \right)}\mp{4}\frac{ i \vartheta _1^{\prime }\left(z,q \right)}{\vartheta _1\left(z,q \right)}-4.
  \label{IdentityIVI:eq4}
\end{equation}

%\paragraph{Identity VII} The series sum 
%\begin{equation}
%  \begin{split}
%    \sum_{n=1}^{\infty} \frac{q^n}{(1-q^n)^2}&=\frac{1}{8}\left(1-\frac{\theta_{2}^{\prime\prime}}{\theta_2}-\frac{\theta_{3}^{\prime\prime}}{\theta_3}\right),\\
%    &=\frac{-1}{8}+\frac{1}{6} E_{2}(2 \tau)-\frac{1}{24} E_{2}(\tau),
%    \end{split}
%  \label{IdentityVII}
%\end{equation}
%has the above two quivalent functional form in terms of either elliptic $\vartheta_2(0;q)$ and $\vartheta_3(0;q)$ or Eisenstein series $E_{2}(\tau)$.

\paragraph{Identity V}
The logarithm of $\vartheta_1(z,q)$ is given by
\begin{equation}
  \log (\vartheta _1(z,q))=4 \sum _{k=1}^{\infty } \frac{q^{2 k} \sin ^2(k z)}{k \left(1-q^{2 k}\right)}+\log (\vartheta _1^{\prime }(0,q))+\log (\sin (z)).
  \label{IdentityV:eq1}  
\end{equation}
  The first derivative of the logarithm of $\vartheta _1(z,q)$ 
\begin{equation}
\frac{\vartheta _1^{\prime }(z,q)}{\vartheta _1(z,q)}=4 \sum _{k=1}^{\infty } \frac{q^{2 k} \sin (2 k z)}{1-q^{2 k}}+\cot(z). 
\label{IdentityV}  
\end{equation}

\paragraph{Identity VI} 
The derivative of the Eq.\eqref{IdentityV:eq1} at $z=0$ blow up owing to the involved two terms. The divergent sum is regularized with $\mathbf{\zeta}$ function. The divergence of the second term is eliminated by expanding $cot(z)$ around $x=0$ such that
\begin{equation}
\begin{split}
\cot(z)&=-\frac{i \left(e^{-i z}+e^{i z}\right)}{e^{-i z}-e^{i z}},\\
&=\frac{x+\frac{1}{x}}{\frac{1}{x}-x},\\
&=1+2 x^2+2 x^4+2 x^6+2 x^8+2 x^{10}+...,\\
&=-i (1 + 2 \zeta(0)),\\
&=0,
\end{split}
\label{IdentityVI:eq1}  
\end{equation}
since $\zeta(0)=-\dfrac{1}{2}$.

Similarly, the derivative of the Eq.\eqref{IdentityV} at $z=0$ produce divergence of $\csc^2(z)$. Let $x=e^{-iz}$ and expanding $\csc(z)^2$ around $x=0$
\begin{equation}
\begin{split}
 \csc(x)^2&=x^2+2 x^4+3  x^6+4 x^8+5 x^{10}+...,\\
 \lim_{x\to 1} \csc(x)^2 &=(1+2+3+4+...)=\sum_{k=1}^{\infty} k,\\
 &=\mathbf{\zeta}(1),\\
 &=-1/12.
\end{split}
\label{IdentityVI:eq2}  
\end{equation}
The second derivative of the Eq.\eqref{IdentityV} at $z=0$ produce divergence of $2\cot(x)\csc(x)^2$. Expanding around $x=0$
\begin{equation}
\begin{split}
 \cot(x)\csc(x)^2&=8i( x^2+4 x^4+9 x^6+16 x^8+25 x^{10}+...),\\
 \lim_{x\to 1} \cot(x)\csc(x)^2 &=8i(1+4+9+16+...)=\sum_{k=1}^{\infty} k^{2},\\
 &=\mathbf{\zeta}(2),\\
 &=0.
\end{split}
\label{IdentityVI:eq3}  
\end{equation}

Expanding the third derivative of the Eq.\eqref{IdentityV} produces $-2 \csc^4(x)-4 \cot^2(x) \csc^2(x)$

\begin{equation}
\begin{split}
-2 \csc^4(x)-4 \cot^2(x) \csc^2(x)&=-16(x^2+8 x^4+27 x^6+64 x^8+125 x^{10} +...),\\
\lim_{x\to 1} \left(-2 \csc^4(x)-4 \cot^2(x) \csc^2(x)\right) &=-16(1+8+27+64+...)\\
 &=-16\sum_{k=1}^{\infty} k^{3},\\
 &=\mathbf{\zeta}(3),\\
 &=\dfrac{-16}{120}.
\end{split}
\label{IdentityVI:eq4}  
\end{equation}

The fourth derivative follow similarly,

\begin{equation}
\begin{split}
8 \cot^3(z) \csc^2(z)+16 \cot(z) \csc^4(z) &=-32i(x^2+16 x^4 + 81 x^6 + 256 x^8 + 625 X^{10}+...),\\
\lim_{x\to 1} \left(8 \cot^3(z) \csc^2(z)+16 \cot(z) \csc^4(z) \right)&=-32i(1+16+81+265+...)\\
&=-16\sum_{k=1}^{\infty} k^{4},\\
 &=\mathbf{\zeta}(4),\\
 &=0 .
\end{split}
\label{IdentityVI:eq3}  
\end{equation}
%\paragraph{Identity X}
%   The following identity (See Ref.~\cite{SHEN1993299}) gives series representation for the quotient of  elliptic $\vartheta$ function and its second derivative
%\begin{equation}
%  \frac{\vartheta_1^{\prime \prime}(z,q)}{\vartheta_1(z,q)} = 16 \sum_{n=1}^{\infty} \frac{q^{n} \cos \left(2 i \pi  n \left(z \right)\right)}{\left(1 -q^{n} \right)^2}+8 \sum _{n=1}^{\infty } \frac{n q^{n} }{1-q^{n}}-1.
%\label{IdentityX}
%\end{equation}

\paragraph{Identity VII} The Eisenstein series is defined as
\begin{equation}
  E_{2k}(\tau)=1+(-1)^k \frac{4k}{B_{k}} \sum_{n=1}^{\infty} \frac{n^{2k-1}q^{n}}{1-q^n},
\label{IdentityVII}
\end{equation}
%=1+(−1)^k \frac{4k}{B_{k}} \sum_{n=1}^{\infty} \frac{n^{2k-1}}{1-q^n}.
where $B_{k}$ are Bernoulli numbers which are given in terms of Riemann zeta functions as 
\begin{equation}
\mathbf{\zeta}(2k)=\frac{B_k (2 \pi)^{k}}{2k!},~~~~~~~~~\mathbf{\zeta}(1-2k)=(-1)^{k}\frac{B_k}{2k}.
\label{eq:Bernoulli_Numbers}	
\end{equation}
for even and odd numbers respectively.\\
In particular $E_{2}$ and $E_{4}$ are given by
\begin{equation}
E_{2}(\tau) = 1 -24 \sum_{n=1}^{\infty} \frac{n q^{n}}{1-q^n}.\\\nonumber
\label{E2}
\end{equation}
and
\begin{equation}
E_{4}(\tau) = 1 +240 \sum_{n=1}^{\infty} \frac{n^3 q^{n}}{1-q^n}.\\\nonumber
\label{E4}
\end{equation}

\paragraph{Identity VIII} The following products appear in Eq.\eqref{S2} as a result of the algebraic manipulations involved in the sums and are related to elliptic $\vartheta_4$ through the identities given below\\
 
 \begin{equation}
   \begin{split}
     \mathcal{P}_1&= \prod _{k=1}^{\infty } \left(e^{-\frac{\pi  k L}{R}-\frac{2 \pi  L}{R}+\frac{2 \pi  t}{R}}-1\right) \left(1-e^{-\frac{\pi  k L}{R}+\frac{\pi  L}{R}-\frac{2 \pi  t}{R}}\right),\\     
     &=\prod _{k=1}^{\infty }(e^{-\frac{\pi  k L}{R}+\frac{\pi  L}{R}-\frac{2 \pi  t}{R}}+e^{-\frac{\pi  k L}{R}-\frac{2 \pi  L}{R}+\frac{2 \pi  t}{R}}-e^{-\frac{2 \pi  k L}{R}-\frac{\pi  L}{R}}-1),\\
     &=\prod _{k=1}^{\infty } (-1-e^{-\frac{\pi  k L}{R}-\frac{\pi  L}{2 R}}+e^{-\frac{1}{2} \left(\frac{2 \pi  k L}{R}+\frac{\pi  L}{R}\right)} \left(e^{\frac{\pi  k L}{R}+\frac{2 \pi  L}{R}-\frac{2 \pi  t}{R}}+e^{\frac{\pi  k L}{R}-\frac{\pi  L}{R}+\frac{2 \pi  t}{R}}\right)/e^{-\frac{1}{2} \left(\frac{2 \pi  k L}{R}+\frac{\pi  L}{R}\right)})\\     
     &=\prod _{k=1}^{\infty }( -1-e^{-\frac{\pi  k L}{R}-\frac{\pi  L}{2 R}}+e^{\frac{1}{2} \left(\frac{2 \pi  k L}{R}+\frac{\pi  L}{R}\right)} \cosh \left(\frac{3 \pi  L}{2 R}-\frac{\pi  t}{R}\right)),\\
     &=\frac{\vartheta _4\left(\frac{i (3 L \pi )}{4 R}-\frac{i (\pi  t)}{ R},e^{-\frac{L \pi }{2 R}}\right)}{G_{1}};~~ \text{with}~~G_{1}=\prod _{k=1}^{\infty } \left(1-e^{-\frac{\pi  k L}{R}}\right).\\
   \end{split}
   \label{IdentityVIII}
 \end{equation}

\paragraph{Identity IX} The modular increase of the argument of the elliptic $\vartheta_4$ is given by the identity 
\begin{equation}
  \vartheta_4(z+i m \log(q),q)=(-1)^m q^{-m^2} e^{2 i m z} \vartheta_4(z,q),
  \label{IdentityIX}
\end{equation}
$\forall m \in \mathcal{Z}$

\paragraph{Identity X} The elliptic $\vartheta_4(z,q)$ are related to $\vartheta_1(z,q)$ functions 
through the relation
\begin{equation}
\vartheta _1(z,q)=i \sqrt[4]{q} e^{-i z} \vartheta _4\left(z+\frac{\pi  \tau }{2},q\right),
\label{IdentityX}
\end{equation}

%\paragraph{Identity XV} The following identity for logarithm of quotient involving $\vartheta_1$ of holds for any $\alpha, \beta \in \mathcal{C}$ 
%\begin{equation}
%  \log \left(\frac{\vartheta _1(\alpha+\beta,q)}{\vartheta _1(\alpha-\beta,q)}\right)=4 \sum _{k=1}^{\infty } \frac{q^{2 k} \sin (2 \alpha k) \sin (2 \beta k)}{k \left(1-q^{2 k}\right)}+\log \left(\frac{\sin (\alpha+\beta)}{\sin (\alpha-\beta)}\right).
%\label{IdentityXV}
%\end{equation}

%\paragraph{Identity XI} The logarithm of quotient of the following $\vartheta _4$ functions in series form is given by 

%\begin{equation}
%  \log\left(\frac{\vartheta _4(z,q)}{\vartheta _4(0,q)}\right)=4 \sum _{n=1}^{\infty } \frac{ q^n \sin^{2} (i n z)}{n\left(1-q^{2 n}\right)}.
%\label{IdentityXI}
%\end{equation}

\paragraph{Identity XI} The logarithmic derivative of $\vartheta _4^{\prime }(z,q)$ in series form is given by 
\begin{equation}
\frac{\vartheta _4^{\prime }(z,q)}{\vartheta _4(z,q)}=4 \sum _{n=1}^{\infty } \frac{q^n \sin (2 n z)}{1-q^{2 n}}.
\label{IdentityXI}
\end{equation}

\paragraph{Identity XII}~\\

Let $\varphi=\frac{\pi  (L_{T}-4 \zeta^{0})}{4 R}$ and
\begin{equation}
\begin{split}
(z_1=x+iy; z_2=-x+iy),\\   
x=\frac{\pi \zeta^0}{R};y=- \frac{\pi \zeta^1}{R},
\end{split}
\end{equation}
the sum of the series 
\begin{equation}
4 \sum _{m=1}^{\infty } \frac{e^{-\frac{\pi  L_{T} m}{2 R}} \left(\sinh \left(2 m \varphi-\frac{2 i \pi  m \zeta^{1}}{R}\right)\right)}{1-e^{-\frac{\pi  L_{T} m}{R}}}=\frac{\vartheta_4'}{\vartheta_4}(z_1+\frac{\pi \tau}{2}),
\label{Sinh}  
\end{equation}
can be expressed in the form of Jacobi elliptic function $\vartheta_{4}(\tau)$ (Eq.\eqref{IdentityIX}-Eq\eqref{IdentityXI}. In addition, the following invariant subtraction follows from Eq.\eqref{IdentityXI} 
 \begin{equation}
   \frac{\vartheta_4'}{\vartheta_4}(z_2+\frac{\pi \tau}{2})-\frac{\vartheta_4'}{\vartheta_4}(z_1+\frac{\pi \tau}{2})=\frac{\vartheta_1'}{\vartheta_1}(z_2,q)-\frac{\vartheta_1'}{\vartheta_1}(z_1,q),
\label{Subtraction}  
 \end{equation}
From both of Eq.\eqref{Sinh} and Eq.\eqref{Subtraction}, the following equivalence relation holds 
 \begin{equation}
   \frac{\vartheta_1'}{\vartheta_1}(z_2+\frac{\pi \tau}{2})-\frac{\vartheta_1'}{\vartheta_1}(z_1+\frac{\pi \tau}{2})=\sum _{m=1}^{\infty } \frac{e^{-\frac{\pi  L_{T} m}{2 R}} \left(\sinh \left(2 m \varphi-\frac{2 i \pi  m \zeta^{1}}{R}\right)-\sinh \left(2 m \varphi-\frac{2 i \pi  m \zeta^{1}}{R}\right) \right)}{1-e^{-\frac{\pi  L_{T} m}{R}}}
   \label{IdentityXII}
\end{equation}

%%%%%%%%%%%%%%%%%%%%%%%%%%%%%%%%%%%%%%
%%%%%%%%%%%%%%%%%%%%%%%%%%%%%%%%%%%%%%

\section{Appendix: Mean-square width $W^{2}_{b_2}$}
  In this appendix we show in detail the evaluation of the correlator 
\begin{equation}
  \langle \mathbf{X}^{2} S_{b_2}\rangle = b_2 \langle  \left( \mathbf{X} \cdot \mathbf{X}\right) \partial_0 \partial_1 \mathbf{X} \cdot \partial_0 \partial_1 \mathbf{X}) \rangle,
\label{Expectation:Sb2}
\end{equation}
which gives the modifications of the mean-square width of open string by virtue of boundary action $S_{b_2}$.

%%%%%%%%%%%%%%%%%%%%%%%%%%%%%%%%%%%%%%
%%%%%%%%%%%%%%%%%%%%%%%%%%%%%%%%%%%%%%
\subsection{The expectation value of  $W^{2}_{1,b2}$}~\\
%---------------------------------------------
The  Wick contraction corresponding to  
\begin{equation}
W^{2}_{1,b2}=-b_2\langle (\partial_{0} \partial_{1}(\mathbf{X}.\mathbf{X}) \partial_{0} \partial_{1}(\mathbf{X}.\mathbf{X})) \rangle,
\end{equation}
upon point splitting, the expectation value of this correlator is given by
\begin{align}
W^2_{1,b_2}=&-(D-2)b_{2} \left\{ \lim_{\zeta^1 \to 0}+\lim_{\zeta^1 \to R} \right\} \int_{\partial \Sigma} d \zeta'_{0} \lim_{\substack{\zeta' \to \zeta }} \Big[ \partial_{0} \partial_{1} G(\mathbf{\zeta;\zeta'})\partial_{0} \partial_{1} G(\mathbf{\zeta';\zeta}) \Big].
\label{eq:Integral}
\end{align}

 The correlator $G (\zeta', \zeta^{\prime}_{0}; \zeta , \zeta_{0} )\equiv  \langle X(\zeta^{0},\zeta^{1}) X(\zeta^{0'}, \zeta^{1'}) \rangle$ on a cylinder of size $RL_{T}$ with fixed boundary conditions at $\zeta^{1} = 0$ and $\zeta^{1} = R$ and periodic boundary conditions in $\zeta^{0}$ with period $L_{T}$ is given by 
%---------------------------------------------
\begin{equation}
\begin{split}
G(\mathbf{\zeta;\zeta'})=  \frac{1}{\pi  \sigma }\sum _{n=1}^{\infty } \frac{1}{n \left(1-q^n\right)}\sin \left(\frac{\pi  n \zeta}{R}\right) \sin \left(\frac{\pi  n \zeta'}{R}\right)  \left(q^n e^{\frac{\pi  n (\zeta^0-\zeta^{0'}}{R}}+e^{-\frac{\pi n (\zeta^0-\zeta^{0'}}{R}}\right).
\end{split}
\label{eq:Gauss}
\end{equation}
The expectation value corresponding to the above correlator 
\begin{equation}
\begin{split}
W^{2}_{1,b_2}=&-\frac{\pi^2 b_2}{R^4 \sigma^2}\left\{ \lim_{\zeta^1 \to 0}+\lim_{\zeta^1 \to R} \right\}\int_{\partial \Sigma}d\zeta^{0} \lim_{\substack{\zeta' \to \zeta }}\sum_{n,m}\frac{mn \pi^2}{(-1+q^m)(-1+q^n)}
\Bigg\{ 2e^{-\frac{\pi (m+n)(\zeta^{0}-\zeta^{0'})}{R}}\\
&\left(-1+e^{\frac{2\pi  m(\zeta^{0}-\zeta^{0'})}{R}} q^m \right)\times(-1 + e^{\frac{2\pi m(\zeta^{0}-\zeta^{0'})}{R}}q^n) 
\cos \left(\frac{\pi n\zeta^{1}}{R} \right) 
\cos \left(\frac{\pi m\zeta^{1'}}{R} \right)\\
&\sin \left(\frac{\pi m\zeta^{1}}{R} \right) 
\sin \left(\frac{\pi n\zeta^{1'}}{R} \right) \Bigg\}.
\end{split}
\end{equation}
Making the substitutions\\
\begin{equation}
\zeta^{1'}= \zeta^{1} +\epsilon;~~~~
\zeta^{0'}= L_{T}-\zeta^{0}-\epsilon';
\end{equation}
the expectation value now reads
\begin{equation}
\begin{split}
  W^{2}_{1,b_2}=&-\frac{2 \pi^{2}(D-2)b_2}{R^4 \sigma^2}\left\{ \lim_{\zeta^1 \to 0}+\lim_{\zeta^1 \to R} \right\}\int_{\partial \Sigma}d\zeta^{0} \lim_{\substack{\epsilon' \to 0 \\ \epsilon \to 0}} \sum_{m=1,n=1} \frac{mn}{(-1+q^m)(-1+q^n)}\\
&  e^{\frac{\pi  (m+n) \left(L_{T}-2\zeta^{0}-\epsilon' \right)}{R}}\big(-1+e^{\frac{2 \pi  m \left(-L_{T}+2\zeta^{0}+\epsilon' \right)}{R}}q^m \big) \big(-1+e^{\frac{2 \pi  n \left(-L_{T}+2\zeta^{0}+\epsilon' \right)}{R}}q^n\big)\\
&\times \cos \left(\frac{\pi  n \zeta^{1}}{R}\right)\cos \left(\frac{\pi  m (\zeta^{1}+\epsilon)}{R}\right) \sin \left(\frac{\pi  m \zeta^{1}}{R}\right) \sin \left(\frac{\pi  n (\zeta^{1}+\epsilon)}{R}\right).\\
\end{split}
\end{equation}
After evaluating the limits $\epsilon \rightarrow 0$ and $\epsilon' \rightarrow 0$
\begin{equation}
\begin{split}  
  W^{2}_{1,b_2}= & -\frac{\pi ^2 b_2 (D-2)}{2 R^4 \sigma^2}\left\{ \lim_{\zeta^1 \to 0}+\lim_{\zeta^1 \to R} \right\} \int_{\partial \Sigma}d\zeta^{0} \sum _{m=1}^{\infty } \frac{m e^{\frac{\pi  m (L_{T}-2 \zeta^{0})}{R}}}{q^m-1} \sin \left(\frac{2 \pi  m \zeta^{1}}{R}\right)\\
&  \left(q^m e^{\frac{2 \pi  m (2 \zeta^{0}-L_{T})}{R}}-1\right) \sum _{n=1}^{\infty } \frac{n e^{\frac{\pi n(L_{T}-2\zeta^{0})}{R}}}{q^n-1} \sin \left(\frac{2 \pi  n \zeta^{1}}{R}\right) \left(q^n e^{\frac{2 \pi  n (2 \zeta^{0}-L_{T})}{R}}-1\right).
\end{split}
\label{1w21b1}
\end{equation}
Let us consider one of the above symmetric sums over $m$ and $n$  
\begin{equation}
\mathcal{S}=\sum _{n=1}^{\infty } \frac{n e^{\frac{\pi  n (L_{T}-2 \zeta^{0})}{R}}}{q^n-1} \sin \left(\frac{2 \pi  n \zeta^{1}}{R}\right) \left(q^n e^{\frac{2 \pi  n (2 \zeta^{0}-L_{T})}{R}}-1\right).
\end{equation}
Expanding the denominator 
\begin{equation}
\begin{split}
\mathcal{S}=\sum_{n=1}^{\infty }\sum_{k=0}^{\infty } n & \Bigg(-\frac{i}{2}  e^{ \left(-\frac{\pi  k L_{T} n}{R}-\frac{2 \pi  L_{T} n}{R}+\frac{2 \pi  n \zeta^0}{R}+\frac{2 i \pi  n \zeta^1}{R}\right)}+\frac{i}{2}  e^{\left(-\frac{\pi  k L_{T} n}{R}+\frac{\pi  L_{T} n}{R}-\frac{2 \pi  n \zeta^0}{R}+\frac{2 i \pi  n \zeta^1}{R}\right)}\\
&+\frac{i}{2}  e^{\left(-\frac{\pi  k L_{T} n}{R}-\frac{2 \pi  L_{T} n}{R}+\frac{2 \pi  n \zeta^0}{R}-\frac{2 i \pi  n \zeta^1}{R}\right)}-\frac{i}{2} e^{ \left(-\frac{\pi  k L_{T} n}{R}+\frac{\pi  L_{T} n}{R}-\frac{2 \pi  n \zeta^0}{R}-\frac{2 i \pi  n \zeta^1}{R}\right)}\Bigg).
\end{split}
\end{equation}
%\begin{equation}
%  \begin{split}
%  \mathcal{S}=\sum_{n=1}^{\infty }\sum_{k=0}^{\infty }&-n \Bigg(\sinh \left(\frac{\pi n  k L}{R}+\frac{2 \pi n  L}{R}-\frac{2 \pi  n \zeta^{0}}{R}\right)- \sinh \left(-\frac{\pi n k L}{R}+\frac{\pi n L}{R}-\frac{2 \pi  n \zeta^{0}}{R}\right)\\
%  &- \cosh \left(\frac{\pi n k L}{R}+\frac{2 \pi n L}{R}-\frac{2 \pi  n \zeta^{0}}{R}\right)+ \cosh \left(-\frac{\pi n k L}{R}+\frac{\pi n L}{R}-\frac{2 \pi  n \zeta^{0}}{R}\right)\Bigg)
%  \end{split}
%\end{equation}
  The sum over $n$ of each term can be put in closed forms in terms of hyperbolic function  
\begin{equation}
\begin{split}
&\mathcal{S}=%\sum_{k=0}^{\infty }\frac{1}{4} \text{csch}^2\left(\frac{-\pi  ((k-1) L-2 \zeta^{0})}{2 R}\right)-\frac{1}{4}\text{csch}^2\left(\frac{-\pi L_{T} (-k-2) }{2 R}-\frac{\pi  \zeta^0}{R}\right)
\frac{i}{8}\sum_{k=0}    \Bigg(-\text{csch}\left(\frac{-\pi L_{T} (-k-2) }{2 R}-\frac{\pi  \zeta^0}{R}-\frac{i \pi  \zeta^1}{R}\right)^{2}- \text{csch}\left(\frac{-\pi L_{T} (k-1) }{2 R}-\frac{\pi  \zeta^0}{R}-\frac{i \pi  \zeta^1}{R}\right)^{2}\\
    &+\text{csch}\left( \frac{-\pi L_{T}(-k-2)}{2 R}-\frac{\pi  \zeta^0}{R}+\frac{i \pi  \zeta^1}{R}\right)^{2}+ \text{csch}\left(\frac{-\pi L_{T} (k-1)}{2 R}-\frac{\pi  \zeta^0}{R}+\frac{i \pi  \zeta^1}{R}\right)^{2}\Bigg).\\
\end{split}
\end{equation}
   The sum index can be redefined $k$ to $k-1$ in the second term of the above equation such that $\sum_{k=0}^{\infty} \rightarrow \sum_{k=1}^{\infty}$, followed by sign inversion to $k$ to $-k$ such that $\sum_{k=1}^{\infty} \rightarrow \sum_{k=-1}^{-\infty}$ the sum reads 
\begin{equation}
\begin{split}
  & \mathcal{S}=\\
  &\frac{i}{8} \Bigg(-\sum _{k=0}^{\infty } \text{csch}\left(\frac{-\pi L_{T} (k-1) }{2 R}-\frac{\pi  \zeta^0}{R}-\frac{i \pi  \zeta^1}{R}\right)^{2}-\sum _{k=-1}^{-\infty } \text{csch}\left(\frac{-\pi  L_{T} (k-1) }{2 R}-\frac{\pi  \zeta^0}{R}-\frac{i \pi  \zeta^1}{R}\right)^{2}\\
  &+\sum _{k=0}^{\infty } \text{csch}\left(\frac{-\pi L_{T} (k-1)}{2 R}-\frac{\pi  \zeta^0}{R}+\frac{i \pi  \zeta^1}{R}\right)^{2}+\sum _{k=-1}^{-\infty } \text{csch}\left( \frac{-\pi L_{T}(k-1) }{2 R}-\frac{\pi  \zeta^0}{R}+\frac{i \pi  \zeta^1}{R}\right)^{2}\Bigg).\\
\end{split}
\end{equation}

  Each two consecutive terms in the above equation form a term of complete sum from $-\infty$ to $\infty$. For a small value of $\zeta^{1}$ the difference between the resultant two terms can be written in the form of differential such that  
\begin{equation}
  \mathcal{S}=\frac{i}{4}\frac{\partial }{\partial \zeta^1}\sum_{k=-\infty}^{\infty } \text{csch}\left(\frac{-\pi L_{T}(k-1) }{2 R}-\frac{\pi  \zeta^0}{R}+\frac{i \pi  \zeta^1}{R}\right)^2 d\zeta^{1}
%&  \frac{-i R}{4\pi}\left(\frac{d U(\zeta)}{d \zeta^{0}}\right).
\end{equation}
%Let
%\begin{equation}
%   \frac{d U(\zeta)}{d \zeta^{0}}= \sum _{k=-\infty}^{\infty } \text{csch}\left(\frac{-\pi L_{T} k}{2 R}-\frac{\pi  \zeta^0}{R}\right)^2
%  -\sum _{k=-\infty}^{\infty } \coth\left(-\frac{\pi L_{T}k}{2 R}+\phi\right)
%\end{equation}

%where $\phi=\left(\frac{\pi  L_{T}}{2 R}-\frac{\pi  \zeta^{0}}{R}-\frac{i \pi  \zeta^{1}}{R}\right)$.
%newpage\left\{ \lim_{\zeta^1 \to 0}+\lim_{\zeta^1 \to R} \right\}
The integral Eq.\eqref{1w21b1} can now be brought into the closed form
  \begin{equation}
    W^{2}_{1,b_2}=\frac{b_2  (D-2)}{32 R^2 \sigma^2}d^{2}\zeta^{1} \int _{\partial \Sigma}  d\zeta^0
\left(\frac{d }{d \zeta^{0}} U(\zeta^{0})\right)^{2}. 
  \label{Integral}
  \end{equation}
with
\begin{equation}
\begin{split}
U(\zeta^{0})&= \frac{\partial }{\partial \zeta^1} \sum_{k=-\infty }^{\infty } \coth \left(\frac{-\pi L_{T}(k-1) }{2 R}-\frac{\pi  \zeta^0}{R}+\frac{i \pi  \zeta^1}{R}  \right)\\
&=\frac{i\partial }{\partial \zeta^1}\frac{ \vartheta _1^{\prime }\left(\frac{i \pi  \zeta^{0}}{R}-\frac{ \pi  \zeta^{1}}{R},e^{-\frac{\pi  L}{2 R}}\right)}{\vartheta _1\left(\frac{i \pi \zeta^{0}}{R}-\frac{ \pi  \zeta^{1}}{R},e^{-\frac{\pi  L}{2 R}}\right)}
\end{split}
\end{equation}
 where we have used Eq.\eqref{IdentityII} to bring the sum into closed form in terms of Jacobi elliptic $\theta$ function.
%  However, the square of the first derivative is related to the second differentiation via,
%\begin{equation}
%\begin{split}
%2  \left(\frac{\partial U(\zeta^{0})}{\partial \zeta^{0}}\right)^2  = \frac{\partial }{\partial \zeta^{0}}\frac{\partial U(\zeta^{0})^2}{\partial \zeta^{0}} -2  U(\zeta^{0}) \frac{\partial }{\partial \zeta^{0}}\frac{\partial U(\zeta^{0})}{\partial \zeta^{0}},
%\end{split}
%\label{ddU}
%\end{equation}

 The above equation Eq.\eqref{Integral} involves an integral over the square of the derivative with respect to $\zeta_{0}$ which is nontrivial to evaluate directly. After integrating by parts 
\begin{equation}
 \int_{0}^{L_{T}} \left(\frac{\partial U(\zeta^{0})}{\partial \zeta^{0}}\right)^2 \, d\zeta^{0}=\left[U(\zeta^{0}) \frac{d U(\zeta^{0})}{d \zeta^{0}}\right]_{0}^{L_{T}}-\int_0^{L_{T}} U(\zeta^{0}) \frac{\partial^2 U(\zeta^{0})}{d \zeta^{0^2}} d\zeta^0,
\label{DoubleInt}
\end{equation}

\begin{figure*}[!hpt]	
	\begin{center}						
		\subfigure[]{\includegraphics[scale=0.9]{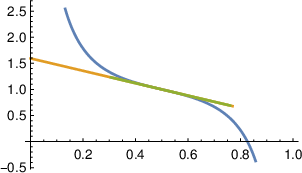}}
		\subfigure[]{\includegraphics[scale=0.53]{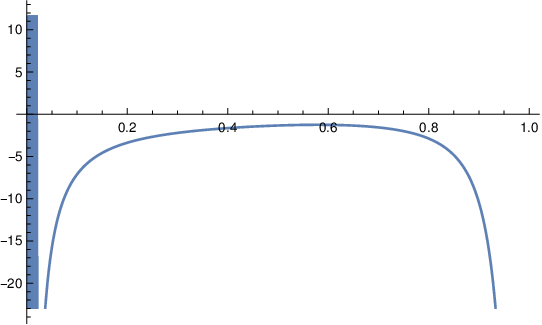}}
                \subfigure[]{\includegraphics[scale=0.53]{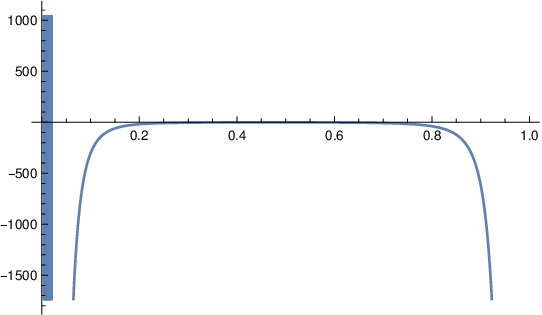}}
		\caption{(a)Compares the logarithmic derivative $U=\frac{\vartheta_1^{\prime }}{\vartheta_{1}}\left(\zeta^{0} ,q \right)$ to its linearized form (b)The derivative $\frac{\partial U(\zeta^{0})}{\partial \zeta^{0}}$ of the logarithmic dervative in the Eq.\eqref{byparts} (c)The contribution of $\frac{\partial^{2} U(\zeta^{0})}{\partial \zeta^{0^2}}$ to the integrand of Eq.~\eqref{byparts} }
		\label{Approx}		
	\end{center}
\end{figure*}
  The second derivative of $\frac{\partial^{2} U(\zeta^{0})}{\partial \zeta^{0^2}}$ in the integrand of the second term diverges much faster than $U(\zeta^{0})$ near the integral limits. It can be a good approximtion to expand  the logarithmic derivative $U(\zeta^{0})$ around $\frac{L_T}{4}$ where it assumes almost a linear form (see Fig.\ref{Approx} for comparison)

\begin{equation}
\begin{split}
  2 \int_0^{L_{T}/2}& U(\zeta^{0})\frac{d^2 U(\zeta^{0})}{d \zeta^{0^2}} d\zeta^0 \\
                =&\Bigg[2U\left(\frac{L_{T}}{4}\right) \frac{\partial U(\zeta^{0})}{\partial \zeta^{0}} \Bigg]_{0}^{L_{T}/2}+ \sum_{n=1}^{\infty }  \left(\frac{i\pi}{R}\right)^{n}U^{(n)}\left(\frac{L_{T}}{4}\right)\int_0^{L_{T}/2} \left(\zeta^{0}-\frac{L_{T}}{4}\right)^{n} \frac{\partial^2 U(\zeta^{0})}{d \zeta^{0^2}} \partial \zeta^0,\\
                =&\Bigg[2U\left(\frac{L_{T}}{4}\right) \frac{\partial U(\zeta^{0})}{\partial \zeta^{0}} \Bigg]_{0}^{L_{T}/2}+ \frac{i\pi}{R}U'\left(\frac{L_{T}}{4}\right)\Bigg[\zeta^{0} \frac{\partial U(\zeta^{0})}{d \zeta^{0}}- \frac{L_{T}}{4} \frac{\partial U(\zeta^{0})}{\partial \zeta^{0}}-U(\zeta^{0}) \Bigg]_{0}^{L_{T}/2} +...
\label{byparts}  
\end{split}
\end{equation}

where the prime in $U^{\prime}(z)$ indicates differentiation with respect to the argument $z$.

The integration limits are evaluated making use of Eq.\eqref{IdentityIV:eq1}, Eq.\eqref{IdentityIV:eq2} and Eq.\eqref{IdentityIV:eq3} of Identity(V) for the modular increase/decrease $\pi\tau$ of the argument of the logarithmic derivatives. The first term in Eq.\eqref{DoubleInt} is

  \begin{equation}
    \begin{split}
\left\{ \lim_{\zeta^1 \to 0}+\lim_{\zeta^1 \to R} \right\}      \left[U(\zeta^{0}) \frac{d U(\zeta^{0})}{d \zeta^{0}}\right]_{0}^{L_{T}} d^{2}\zeta^{1}&=\frac{-4\pi}{R}\Bigg[ \frac{\vartheta _1^{\prime }}{\vartheta_{1}}\left(0 ,q \right)^{'} \frac{\vartheta _1^{\prime }}{\vartheta_{1}}\left(0 ,q \right)^{''}\Bigg]d^{2}\zeta^{1},\\
%      -i\frac{\vartheta _1^{\prime }}{\vartheta_{1}}\left(0 ,q \right)\Bigg(\frac{\vartheta _1^{\prime }}{\vartheta_{1}}\left(0 ,q \right)\Bigg)^{'}\Bigg)\\
%      &=\frac{-4\pi}{R}\Bigg(\frac{\vartheta _1^{\prime }}{\vartheta_{1}}\left(0 ,q \right)\Bigg)^{'}.
    \end{split}
    \label{First}
  \end{equation}
  In the above equation, one should take into account the sign flips owing to the discontinuity jumps near the integration limits.

  The second term reads
\begin{equation}
\begin{split}
  -2\left\{ \lim_{\zeta^1 \to 0}+\lim_{\zeta^1 \to R} \right\}d^{2}\zeta^{1} \int_0^{L_{T}/2} U(\zeta^{0})&\frac{d^2 U(\zeta^{0})}{d \zeta^{0^2}} d\zeta^0\\
  %\frac{-\pi}{R}U^{'}\left(\frac{L_{T}}{4}\right)\Bigg[ L_{T} \Bigg(\frac{\vartheta _1^{\prime }}{\vartheta_{1}}\left(0 ,q \right)\Bigg)^{'}+[2+i \frac{\vartheta _1^{\prime }}{\vartheta_{1}}\left(0 ,q \right)]-i\frac{\vartheta_1^{\prime }}{\vartheta_{1}}\left(0 ,q \right)\Bigg] \\
  & =-2\bigg[U\left(\frac{L_{T}}{4}\right)\Bigg(\frac{\vartheta_1^{\prime }}{\vartheta_{1}}\left(0 ,q \right)\Bigg)^{''}+..... \bigg]d^{2}\zeta^{1}.
\end{split}
\label{Second}
\end{equation}
%since the derivatives $U^{(n)}\left(\frac{L_{T}}{4}\right)\simeq 0$ involved in the next to leading order terms of the taylor expansion Eq.\eqref{byparts} are almost vanishing, the outcome divergent sum, $\sum_{n=1}^{\infty}\epsilon$, in Eq.\eqref{Second} is regularized by zeta function yielding $\zeta(\infty)=1$.
%The correlator is cast to the form
%\begin{equation}
%    W^{2}_{1,b_2}=\frac{b_2 (D-2)\pi}{512  R^3 \sigma^2}  \Bigg( \frac{\vartheta _1^{\prime }}{\vartheta_{1}}\left(\frac{0 ,q \right)\Bigg)^{'}.\\
%  \end{split}
%\end{equation}
%The expectation value after collecting the two terms Eq.~\eqref{First} and Eq.~\eqref{Second} is
%\begin{equation}
%  W^{2}_{1,b_2}= \frac{b_2 (D-2)}{32 R^2 \sigma^2}\Bigg[ \frac{-4\pi}{R}\Bigg(\frac{\vartheta _1^{\prime }}{\vartheta_{1}}\left(0 ,q \right)\Bigg)^{'}+\frac{-i\pi}{R}U^{'}\left(\frac{L_{T}}{4}\right)\bigg[\frac{\pi L_{T}}{2 R}\Bigg(\frac{\vartheta_1^{\prime }}{\vartheta_{1}}\left(0 ,q \right)\Bigg)^{'}+2 \bigg]\Bigg]. 
%\end{equation}

The divergent part of the differentiations of the logarithmic derivative at the poles $\zeta^1=0$ can be regularized with $\mathbf{\zeta}$-function using Identities Eq.\ref{IdentityVI:eq1}, Eq.\ref{IdentityVI:eq2}  such that
\begin{equation}
\begin{split}   
\frac{\vartheta^{\prime}_{1}}{\vartheta_{1}}\left(0,q\right)^{\prime}&=\left(4\sum_{n=1}^{\infty}\frac{2n q^{2n}}{1-q^{2n}}-\text{csc}^{2}(\epsilon)\right),\\
&=  \left( \frac{8}{24}\left(1-E_{2}(2\tau)\right)+\frac{1}{12} \right).
\end{split}  
\label{Reg1}
\end{equation}
 The second derivative
\begin{equation}
\begin{split}   
\frac{\vartheta^{\prime}_{1}}{\vartheta_{1}}\left(0,q\right)^{\prime\prime}&=\left(\sum_{n=1}^{\infty}\epsilon-\sum_{n=1}^{\infty}n^2\right),\\
&= \zeta(\infty)+\zeta(-2),\\
&=1\quad .
\end{split}
\label{Reg2}
\end{equation}

% The expectation value finally turn out to be
The expectation value after collecting the two terms Eq.~\eqref{First} and Eq.~\eqref{Second}, parameter redefinition $b_2 ~d^2\zeta^{1} \longrightarrow b_2$ and making use of the regularization Eq.~\eqref{Reg1} and Eq.~\eqref{Reg2} becomes
 \begin{equation}
%   \begin{split}
   W^{2}_{1,b_2}= \frac{-\pi^{3}  b_2 (D-2)}{16 R^5 \sigma^2}\Bigg[-\frac{1}{3}E_{2}(2\tau)-2\frac{\vartheta _1^{\prime }}{\vartheta _1}\left(\frac{\tau}{2},q\right)^{(1)}+\frac{5}{6}\Bigg],
%  \end{split}
\end{equation}
% with
% \begin{equation}
%  \begin{split} 
%    h\left(\frac{\tau}{2}\right)
%    &=\frac{-i}{4}U^{'}\left(\frac{L_{T}}{4}\right)&
%    =\frac{1}{4} \frac{\vartheta^{\prime}_{1}}{\vartheta_{1}}\left(\frac{\tau}{2},q\right)^{\prime}
%    = -2\left[\frac{\vartheta _1^{\prime }}{\vartheta _1}\left(\frac{\tau}{2},q\right)^{(1)}+\frac{\vartheta _1^{\prime }}{\vartheta _1}\left(\frac{\tau}{2},q\right)-\frac{1}{24}\right].
%      -\frac{\vartheta _1^{\prime }}{\vartheta _1}\left(\frac{\tau}{2},q\right)^2+1\right].
%  \end{split}
%  \label{h}
% \end{equation}

%%%%%%%%%%%%%%%%%%%%%%%%%%%%%%%%%%%%%%
%%%%%%%%%%%%%%%%%%%%%%%%%%%%%%%%%%%%%%
\subsection{The expectation value of $W^{2}_{2,b_2}$}~\\
%  $\langle (\mathbf{X}.\mathbf{X}) \partial_{0}^{2} \partial_{1}^{2} (\mathbf{X}.\mathbf{X})\rangle$}~\\
The next possible Wick-contraction of \eqref{Expectation:Sb2} yields the correlator
\begin{equation}
W^{2}_{2,b_2}=-b_2\langle (\mathbf{X}.\mathbf{X}) \partial_{0}^{2} \partial_{1}^{2} (\mathbf{X}.\mathbf{X}) \rangle.
\label{Corb22}  
\end{equation}
The integral over Green function after point splitting of the correlator  
\begin{equation}
W^2_{2,b_2}=-(D-2)b_{2}\left\{ \lim_{\zeta^1 \to 0}+\lim_{\zeta^1 \to R} \right\} \int_{\partial \Sigma} d \zeta'_{0} \lim_{\zeta' \to \zeta} \Big[ G(\mathbf{\zeta;\zeta'}) \partial_{0}^{2} \partial_{1}^{2}G(\mathbf{\zeta';\zeta}) \Big].
\label{Bw2b2}
\end{equation}
Substituting the Green function corresponding to the free propagator \eqref{eq:Gauss} into above \eqref{Bw2b2}
\begin{equation}
\begin{split}
W^{2}_{2,b_2} =\frac{-\pi^2 b_2(D-2)}{R^4 \sigma^2}&\left\{ \lim_{\zeta^1 \to 0}+\lim_{\zeta^1 \to R} \right\}\int_{\partial \Sigma}\lim_{\zeta' \to \zeta} d\zeta^{0}  \sum_{m,n=1}^{\infty}\frac{m^3}{n \left(q^m-1\right) \left(q^n-1\right)}\\
&\times 
\cos\left(\frac{\pi m\zeta^{1}}{R}\right)
\sin\left(\frac{\pi n\zeta^{1}}{R}\right)
\sin\left(\frac{\pi n\zeta^{1'}}{R}\right)
\cos\left(\frac{\pi m\zeta^{1'}}{R}\right)\\
&\times e^{-\frac{\pi (m+n)\left(\zeta^{0}-\zeta^{0'}\right)}{R}} \left(q^m e^{\frac{2 \pi  m \left(\zeta^{0}-\zeta^{0'}\right)}{R}}+1\right)\left(q^n e^{\frac{2\pi  n\left(\zeta^{0}-\zeta^{0'}\right)}{R}}+1\right). 
\end{split} 
\end{equation}
With the substitutions
\begin{equation}
\zeta^{1'}= \zeta^{1} +\epsilon;
\zeta^{0'}= L_{T}-\zeta^{0}-\epsilon';
\end{equation}
the expectation value becomes
\begin{equation}
\begin{split}
  W^{2}_{2,b_2}& =\frac{-\pi^2 b_2 (D-2)}{R^4 \sigma^2} \left\{ \lim_{\zeta^1 \to 0}+\lim_{\zeta^1 \to R} \right\}  \int_{\partial \Sigma} \lim_{\substack{\epsilon' \to 0 \\ \epsilon \to 0}} d\zeta^{0}\Bigg\{\sum_{m,n=1}^{\infty}e^{\frac{\pi (m+n) (L_{T}-2 t-\epsilon' )}{R}}
 \\
  &\times \dfrac{m^3}{n \left(q^m-1\right) \left(q^n-1 \right)}\left(q^n e^{\frac{2\pi n(-L_{T}+2 \zeta^{0}+\epsilon' )}{R}}+1\right)  \left(q^m e^{\frac{2\pi m(-L_{T}+2 \zeta^{0}+\epsilon' )}{R}}+1\right) \\
  &\times  \cos \left(\frac{\pi  m \zeta^{1}}{R}\right) \sin \left(\frac{\pi n\zeta^{1}}{R}\right) \cos \left(\frac{\pi m(\zeta^{1}+\epsilon )}{R}\right) \sin \left(\frac{\pi n(\zeta^{1}+\epsilon )}{R}\right)\Bigg\}.
  %\dfrac{1} {n \left(q^m-1\right) \left(q^n-1\right)}   
\end{split}
\end{equation}
Taking the limit $\epsilon \to 0$, $\epsilon' \to 0$ the expectation value in compact form reads 
\begin{equation}
\begin{split}
W^{2}_{2,b_2} =-\frac{\pi^2 b_2 (D-2)}{R^4 \sigma^2}\left\{ \lim_{\zeta^1 \to 0}+\lim_{\zeta^1 \to R} \right\} \int_{\partial \Sigma} d\zeta^{0} \left( \mathcal{S}_1 \mathcal{S}_2 \right),
\end{split}
\end{equation}
with
\begin{equation}
  \mathcal{S}_1=  \sum _{m=1}^{\infty } m^3 \text{csch} \left(\frac{\pi  L_{T} m}{2R}\right) \cos ^2\left(\frac{\pi m\zeta^{1}}{R}\right) \cosh \left(\frac{\pi m(3L_{T}-4\zeta^{0})}{2 R}\right),\\
  \label{S1}
\end{equation}
and
\begin{equation}
  \mathcal{S}_2= \sum_{n=1}^{\infty} \frac{1}{n} \text{csch}\left(\frac{\pi  L_{T} n}{2R}\right) \sin ^2\left(\frac{\pi n\zeta^{1}}{R}\right) \cosh \left(\frac{\pi n(3 L_{T}-4\zeta^{0})}{2R}\right).
\label{S2}
\end{equation}

  Let us consider each sum in the multiplication separately. Plugging the series expansion of the hyperbolic function 
\begin{equation} 
\text{csch}\left(\frac{\pi  L m}{2 R}\right)=-2 \sum_{k=0}^{\infty } e^{\frac{(2 k+1) (\pi  L_{T} m)}{2 R}},
\end{equation}

into \eqref{S1} for the first sum $S_1$
\begin{equation}
\mathcal{S}_1= \left(\sum _{k=0}^{\infty } \sum _{m=1}^{\infty } (\mathcal{S}_{1,1}(m,k)+\mathcal{S}_{1,2}(m,k))\right),
\end{equation}
such that 
\begin{equation}
  \begin{split}
\mathcal{S}_{1,1}(m,k)&=m^3 \Bigg[-\frac{1}{2} \cosh \left(-\frac{\pi kL_{T}m}{R}+n \pi \alpha_1  \right)-\frac{1}{2} \cosh \left(\frac{\pi  k L_{T} m}{R}+n \pi \alpha_2 \right)\\
&-\frac{1}{4} \cosh \left(-\frac{\pi  kL_{T}m}{R}+n \pi \beta_1 \right)-\frac{1}{4} \cosh \left(\frac{\pi kL_{T}m}{R}+n \pi \beta_2 \right) \\
&-\frac{1}{4} \cosh \left(-\frac{\pi kL_{T}m}{R}+n \pi \beta_1^{*} \right)-\frac{1}{4} \cosh \left(\frac{\pi kL_{T}m}{R}+n \pi \beta_2^{*} \right)\Bigg],
  \end{split}
  \label{S11}
\end{equation}
and
\begin{equation}
  \begin{split}
\mathcal{S}_{1,2}(m,k)&=m^3  \Bigg[-\frac{1}{2} \sinh \left(\frac{\pi kL_{T}m}{R}+n \pi \alpha_2  \right)+\frac{1}{2} \sinh \left(-\frac{\pi  k L_{T} m}{R}+n \pi \alpha_1 \right)\\
&+\frac{1}{4} \sinh \left(\frac{\pi kL_{T}m}{R}+n \pi \beta_2^{*} \right)+\frac{1}{4} \sinh \left(-\frac{\pi kL_{T}m}{R}+n \pi \beta_1^{*}  \right)\\
&-\frac{1}{4} \sinh \left(\frac{\pi  kL_{T}m}{R}+n \pi \beta_2 \right)+\frac{1}{4} \sinh \left(-\frac{\pi kL_{T}m}{R}+n \pi \beta_1 \right) \Bigg],
\end{split}
\label{S12}
\end{equation}
with
\begin{equation}
\begin{split}
  \alpha_1&=\frac{L_{T}}{R}-\frac{\zeta^{0}}{R}  ,\alpha_2=\frac{2 L_{T}}{R}-\frac{2 \zeta^{0}}{R},\\
  \beta_1&= \frac{2 L_{T}}{R}-\frac{2 \zeta^{0}}{R}-\frac{2i \zeta^{1}}{R}, \beta_2= \frac{L_{T}}{R}-\frac{2\zeta^{0}}{R}-\frac{2i  \zeta^{1}}{R}.
\end{split}
\end{equation}
The sum of terms in the above matrices $\mathcal{S}_{1,1}(m,k)=0$ and $\mathcal{S}_{1,2}(m,k)=0$ vanishes term by term $\forall (m, k) $ in the sum.

  The regularization of the first series in the sum proceeds as follows

\begin{equation}
  \begin{split}
 \sum _{k=0}^{\infty } \sum _{m=1}^{\infty } \mathcal{S}_{1,1}(m,k)&=\sum _{k=0}^{\infty } \sum _{m=1}^{\infty } m^3 \cosh \left(\frac{\pi m  (2 k+1) L_{T}}{2 R}\right) \Bigg(\cosh \left(\frac{\pi  m (3 L_{T}-4 \zeta^0+4 i \zeta^1)}{2 R}\right)\\
 &+\cosh \left(\frac{\pi  m (3 L_{T}-4 \zeta^0)}{2 R}\right)\Bigg),\\
  \end{split}
\end{equation}

  where the sum over the running $k$ produces divergent sum over zero
\begin{equation}
\begin{split}
&\sum_{k=0}^{\infty} \sum _{m=1}^{\infty } \mathcal{S}_{1,1}(m,k)=\\
&\sum_{m=1}^{\infty } m^3 \Bigg( \frac{1}{4} \text{csch}\left(\frac{\pi  L m}{2 R}\right)\Bigg(\cosh \left(\frac{\pi  m (3 L_{T}-4 \zeta^0+4 i \zeta^1)}{2 R}\right)+\cosh \left(\frac{\pi  m (3 L_{T}-4 \zeta^0)}{2 R}\right)\Bigg)\\
&-\frac{1}{4} \text{csch} \left(\frac{\pi  L m}{2 R}\right) \Bigg)\Bigg(\cosh \left(\frac{\pi  m (3 L_{T}-4 \zeta^0+4 i \zeta^1)}{2 R}\right)+\cosh \left(\frac{\pi  m (3 L_{T}-4 \zeta^0)}{2 R}\right)\Bigg),
\end{split}
\end{equation}

that is,
\begin{equation}
\begin{split}
\mathcal{S}_1(m^3,k)]=& \left(\sum _{k=0}^{\infty } \sum _{m=1}^{\infty } (\mathcal{S}_{1,1}(m,k)+\mathcal{S}_{1,2}(m,k))\right),\\
=& \zeta(\infty),\\
=& 1.
\end{split}
\label{S1sum}
\end{equation}

Similarly the series $\mathcal{S}_2$,
\begin{equation}
 \mathcal{S}_2=\left(\sum _{k=0}^{\infty } \sum _{n=1}^{\infty }(\mathcal{S}_{2,1}(n,k)+\mathcal{S}_{2,2}(n,k))\right),
  \label{S2sum}
\end{equation}
  corresponds to the sum of the two terms. The sum over the index $n$ can be encapsulated in logarithmic form, the above two series become

\begin{equation}
\begin{split}
\sum_{n} \mathcal{S}_{2,1}(n,k)&=\frac{1}{4} \log \left(2-2 \cosh \left(\frac{-\pi k L_{T}}{R}+\alpha_1 \right)\right) +\frac{1}{4} \log \left(2-2 \cosh \left(\frac{\pi k L_{T}}{R}+\alpha_2 \right)\right)\\
&-\frac{1}{8} \log \left(2-2 \cosh \left(\frac{-\pi k L_{T}}{R}+ \beta_1 \right)\right)-\frac{1}{8} \log \left(2-2 \cosh \left(\frac{\pi k L_{T}}{R}+\beta_2 \right)\right)\\
&-\frac{1}{8} \log \left(2-2 \cosh \left(\frac{\pi k L_{T}}{R}-\beta_1 \right)\right) -\frac{1}{8} \log \left(2-2 \cosh \left(\frac{\pi k L_{T}}{R}+\beta_2^{*} \right)\right)\\
\end{split}
\end{equation}
and

\begin{equation}
  \begin{split}
\sum_{n}\mathcal{S}_{2,2}(n,k)&=-\frac{1}{4} \log \left(-e^{\frac{-\pi k L_{T}}{R}+\alpha_1}   \right)-\frac{1}{4} \log \left(-e^{ -\frac{\pi k L_{T}}{R}-\alpha_2 }\right)+\frac{1}{8} \log \left(-e^{ \frac{-\pi k L_{T}}{R}+\beta_1 }\right)\\ 
&-\frac{1}{8} \log \left(-e^{\frac{\pi k L_{T}}{R}+\beta_2}\right)-\frac{1}{8} \log \left(-e^{\frac{\pi k L_{T}}{R}-\beta_1 } \right)+\frac{1}{8} \log \left(-e^{\frac{-\pi k L_{T}}{R}-\beta_2^{*}  }\right)\\
\end{split}
\end{equation}

The exponential form of the product over the hyperbolic functions,  
\begin{equation}
  \begin{split}
    S_2=& -\frac{1}{4} \log \bigg(-\frac{1}{16}\prod _{k=0}^{\infty } \left(-1+e^{-\frac{\pi  ((k-1) L_{T}+2 (\zeta^{0}-i \zeta^{1}))}{R}}\right) \left(-1+e^{-\frac{\pi  (-k L_{T}+L_{T}-2 (\zeta^{0}+i \zeta^{1}))}{R}}\right)\\
  & \times \left(-1+e^{-\frac{\pi  (-(k+2) L_{T}+2 (\zeta^{0}+i \zeta^{1}))}{R}}\right) \left(-1+e^{-\frac{\pi  ((k+2) L_{T}-2 \zeta^{0}+2 i \zeta^{1})}{R}}\right)\bigg)\\
  & +\sum_{k=0}^{\infty}\frac{\pi  (2 k+1) L_{T}}{ 4 R}+\frac{1}{2} \log\left( \frac{1}{4} \prod _{k=0}^{\infty }\left(e^{-\frac{\pi  k L}{R}-\frac{2 \pi  L}{R}+\frac{2 \pi  t}{R}}-1\right) \left(1-e^{-\frac{\pi  k L}{R}+\frac{\pi  L}{R}-\frac{2 \pi  t}{R}}\right)\right)
    \label{Log}
  \end{split}
\end{equation}
   enables to find a closed form using elliptic $\vartheta_4$ defined by Eq.\eqref{theta4}. Using Eq.\eqref{IdentityVIII} the infinite product within each logarithmic term in the above Eq.\eqref{Log} is pinned down to

  \begin{equation}
    \begin{split}
  S_2&=  \frac{1}{4} \log \left(\frac{ \vartheta _4\left(\frac{i (L_{T} \pi )}{4 R}-\frac{i (\pi  \zeta^{0})}{2 R},e^{-\frac{L_{T} \pi }{2 R}}\right)^2}{\vartheta _4\left(\frac{i \pi  L_{T}}{4 R}+\frac{\pi  \zeta^{1}}{R}-\frac{i \pi  \zeta^{0}}{R},e^{-\frac{L_{T} \pi }{2 R}}\right) \vartheta _4\left(\frac{i \pi  L_{T}}{4 R}-\frac{i \pi  \zeta^{0}}{R}-\frac{\pi  \zeta^{1}}{R},e^{-\frac{L_{T} \pi }{2 R}}\right)}\right).
  \label{S2theta4}
  \end{split}
\end{equation}
  Transforming $\vartheta_{4}(z,q)$ into $\vartheta_{1}(z,q)$ (Eq.\eqref{IdentityIX},Eq.\eqref{IdentityX}) in \eqref{S2theta4} then we have
\begin{equation}
  \begin{split}
    W^{2}_{2,b_2}&=-\frac{\pi ^2 b_2 (D-2)}{4 R^4 \sigma^2}\left\{ \lim_{\zeta^1 \to 0}+\lim_{\zeta^1 \to R} \right\}\\
    &\int_{\partial \Sigma} d\zeta^{0} \log \left(\frac{\vartheta _1\left(\frac{i (\pi  \zeta^{0})}{ R},e^{-\frac{L_{T} \pi }{2 R}}\right)^2}{\vartheta _1\left(\frac{i \pi  \zeta^{0}}{R}-\frac{\pi  \zeta^{1}}{R},e^{-\frac{L_{T} \pi }{2 R}}\right) \vartheta _1\left(\frac{i \pi  \zeta^{0}}{R}+\frac{\pi  \zeta^{1}}{R},e^{-\frac{L_{T} \pi }{2 R}}\right)}\right)
  \end{split}
  \label{IntSplit}
\end{equation}

%The integral over the second term can be manipulated further with the use of Eq.\eqref{IdentityXV}-Identity (XV)-for the logarithm of the quotient of the two elliptic $\vartheta_1$ functions
Rearranging the logarithmic terms in the above equation

%  \begin{equation}
%\begin{split}
%  W^{2}_{2,b_2}& =\frac{\pi ^2 b_2 (D-2)}{4 R^4 \sigma^2}\left\{ \lim_{\zeta^1 \to 0}+\lim_{\zeta^1 \to R} \right\}\\
%  & \times \int _0^{L_{T}}d\zeta^{0}\left[ \log \left(\frac{\vartheta _1\left(\frac{i \pi  \zeta^{0}}{R}+\frac{\pi  \zeta^{1}}{R},e^{-\frac{L_{T} \pi }{2 R}}\right)}{\vartheta _1\left(\frac{i \pi  \zeta^{0}}{ R},e^{-\frac{L_{T} \pi }{2 R}}\right)}\right)+\log \left(\frac{\vartheta _1\left(\frac{i \pi  \zeta^{0}}{R}-\frac{\pi  \zeta^{1}}{R},e^{-\frac{L_{T} \pi }{2 R}}\right)}{\vartheta _1\left(\frac{i \pi  \zeta^{0}}{2 R},e^{-\frac{L_{T} \pi }{2 R}}\right)}\right)\right],\\
%\end{split}
%  \end{equation}
 
  \begin{equation}
   \begin{split}
  W^{2}_{2,b_2}& =\frac{\pi ^2 b_2 (D-2)}{4 R^4 \sigma^2}\times \int _0^{L_{T}}d\zeta^{0} \left\{\lim_{\zeta^1 \to 0}+\lim_{\zeta^1 \to R} \right\} \dfrac{1}{\Delta^{2} \zeta^{1}}\Bigg[ \log \left(\vartheta _1\left(\frac{i \pi  \zeta^{0}}{R}+\frac{\pi  \zeta^{1}}{R},e^{-\frac{L_{T} \pi }{2 R}}\right)\right)\\
  & -2\log\left(\vartheta _1\left(\frac{i \pi  \zeta^{0}}{ R},e^{-\frac{L_{T} \pi }{2 R}} \right)\right) +\log \left(\vartheta _1\left(\frac{i \pi  \zeta^{0}}{R}-\frac{\pi  \zeta^{1}}{R},e^{-\frac{L_{T} \pi }{2 R}}\right)\right)\Bigg]\Delta^{2}\zeta^{1},\\
\end{split}
  \end{equation}

 The difference between the Logarithms in the above equation can be rewritten as a the second derivative  times the second differential $d^{2}\zeta^{1}$ 
\begin{equation}    
  W^{2}_{2,b_2}  =\frac{\pi^3 b_2 (D-2)}{2 R^5 \sigma^2} d^2\zeta^{1}\left[\dfrac{\partial}{\partial \zeta^{1}}\dfrac{\partial}{\partial \zeta^{1}} \int _0^{L_{T}}d\zeta^{0} \log \left( \vartheta _1\left(\frac{i \pi  \zeta^{0}}{R}+\frac{\pi  \zeta^{1}}{R},e^{-\frac{L_{T} \pi }{2 R}}\right)\right)\right]_{\zeta^{1}=0}~~~~~ 
%  &=\frac{\pi ^2 b_2 (D-2)}{4 R^4 \sigma^2}\left\{ \lim_{\zeta^1 \to 0}+\lim_{\zeta^1 \to R}\right\} \Bigg[\frac{-1}{4} \int_0^{L_{T}} \log \left(\text{csch}^2\left(\frac{\pi  \zeta^{0}}{R}\right) \sin^2\left(\frac{\pi  \zeta^{1}}{R}\right)+1\right)d\zeta^{0}\\
%  &-2\sum _{k=1}^{\infty } \int_0^{L_{T}} \frac{e^{-\frac{k (\pi  L)}{R}} \cosh \left(\frac{2 \pi  k \zeta^{0}}{R}\right) \sin^2\left(\frac{\pi k \zeta^{1}}{R}\right)}{k \left(1-e^{-\frac{k (\pi  L)}{R}}\right)}\right)\right)d\zeta^{0}\Bigg].
%\end{split}
\end{equation}
In the limit of small neigborhood of the boundary $d^{2}\zeta^{1}$, the expectation value after parameter redefinition $b_2 d\zeta^{1}\longrightarrow b_2$ and integrating over $\zeta^{0}$ 

\begin{equation}
%  \begin{split}
    W^{2}_{2,b_2} =-\frac{\pi^{3} b_2 (D-2)}{2 R^5 \sigma^2} \left[\frac{\vartheta _1^{\prime}\left(\frac{i \pi \zeta^{0}}{R},e^{-\frac{L_{T} \pi }{2 R}}\right)}{\vartheta _1\left(\frac{i \pi  \zeta^{0}}{2 R},e^{-\frac{L_{T} \pi }{2 R}}\right)}\right]_{0}^{L_{T}}.
%    &=-\frac{\pi^{4} b_2 (D-2)}{4 R^5 \sigma^2}
%    \left\{ \lim_{\zeta^1 \to 0}+\lim_{\zeta^1 \to R} \right\} \sum _{k=1}^{\infty } \frac{e^{-\frac{k (\pi  L)}{R}}  \sinh \left(\frac{2 \pi  k L}{R}\right) \sin ^2\left(\frac{\pi  k \zeta^{1}}{R}\right)}{  k^2 \left(1-e^{-\frac{k (\pi  L)}{R}}\right)}
%\end{split}
\label{I11}
\end{equation}

%where we have used the periodicity property of the logarithmic derivative Eq.\eqref{IdentityVI:eq3}.
After the regularization of the logarithmic derivative using Identities Eq.\ref{IdentityVI:eq1}
\begin{equation}
  \begin{split}
    W^{2}_{2,b_2}& =-\frac{\pi^{3} b_2 (D-2)}{ R^5 \sigma^2}.
 %   \left(\frac{\vartheta _1^{\prime}\left(i\tau/2,e^{-\frac{L_{T} \pi }{2 R}}\right)}{\vartheta _1\left(\frac{i \pi  \zeta^{0}}{2 R},e^{-\frac{L_{T} \pi }{2 R}}\right)}-2\right)\\
%    \left\{ \lim_{\zeta^1 \to 0}+\lim_{\zeta^1 \to R} \right\} \sum _{k=1}^{\infty } \frac{e^{-\frac{k (\pi  L)}{R}}  \sinh \left(\frac{2 \pi  k L}{R}\right) \sin ^2\left(\frac{\pi  k \zeta^{1}}{R}\right)}{  k^2 \left(1-e^{-\frac{k (\pi  L)}{R}}\right)}
\end{split}
\label{I11}
\end{equation}
\section{Appendix(C): Mean-square width $W^{2}_{b_4}$}

  The next non-vanishing Lorentz invariant boundary term has a width expectation value given by
\begin{equation}
\begin{split}
  W^{2}_{b_4}&=\langle \mathbf{X}^{2} S_{b_4}\rangle\\
  &=- b_4 \langle  \left(\mathbf{X} \cdot \mathbf{X}\right) \partial_0^{2} \partial_1 \mathbf{X} \cdot  \partial_0^{2} \partial_1 \mathbf{X}) \rangle.
\end{split}
\label{W2b4total}  
\end{equation}

This expectation value has two Wick contractions that we work out separately in detail in the following.

%%%%%%%%%%%%%%%%%%%%%%%%%%%%%%%%%%%%%%
%%%%%%%%%%%%%%%%%%%%%%%%%%%%%%%%%%%%%%
 \subsection{The expectation value of  $W^{4}_{1,b_4}$}~\\
   The Wick contraction assumes the form
\begin{equation}
 W^{2}_{1,b_4}=-b_4 \langle \partial_{0}^2 \partial_{1} (\mathbf{X} \cdot \mathbf{X}) \partial_{0}^2 \partial_{1} (\mathbf{X} \cdot \mathbf{X}) \rangle
\end{equation}
After point-splitting,
\begin{equation}
\begin{split}
\zeta^{1'}=\zeta^{1}+\epsilon,\\
\zeta^{0'}=L_{T}-\zeta^{0}-\epsilon',
\end{split}
\end{equation}
the correlator in terms of Green function 
\begin{equation}
  W^{2}_{1,b_4}=\left\{ \lim_{\zeta^1 \to 0}+\lim_{\zeta^1 \to R} \right\}\int _{\partial \Sigma}  \partial _{0}^2\partial_{1} G\left(\zeta^0,\zeta^1;\zeta^{0'},\zeta^{1'}\right)\partial _{0'}^2\partial_{1'} G\left(\zeta^0,\zeta^1;\zeta^{0'},\zeta^{1'}\right)d\zeta^0, 
\label{b24}  
\end{equation}

Substituting the free propagator Eq.\eqref{GreenPropagator} into Eq.\eqref{b24} and taking the limits the expectation value would be
\begin{equation}
  \begin{split}
    W^{2}_{1,b_4}&=-\frac{\pi^4 b_4(D-2)}{ 4 R^6 \sigma^2}\left\{ \lim_{\zeta^1 \to 0}+\lim_{\zeta^1 \to R} \right\} \int _{\partial \Sigma} d\zeta^0\Bigg[ \sum_{m}\frac{m^2}{\left(q^m-1\right)} \sin \left(\frac{2 \pi  m \zeta^{1}}{R}\right)\\
    &\bigg( q^m e^{\frac{\pi  m (L_{T}-2 \zeta^{0})}{R}}+e^{-\frac{\pi  m (L_{T}-2 \zeta^{0})}{R}}\bigg) \Bigg]^{2}
%\sum_{n}\frac{1}{ \left(q^n-1\right)}\left(n^2 \sin \left(\frac{2 \pi  n \zeta^{1}}{R}\right) \left(q^n e^{-\frac{\pi  n (L_{T}-2 \zeta^{0})}{R}}+e^{\frac{\pi  n (L_{T}-2 \zeta^{0})}{R}}\right)\right),
  \label{W22b4}  
  \end{split}
\end{equation}
  This results in an integrand of the square of the series 
\begin{equation}
\begin{split}
  \mathcal{S}&=\sum _{m=1}^{\infty } \frac{m^2 \sin \left(\frac{2 \pi  m \zeta^{1}}{R}\right) \left(q^m e^{\frac{\pi  m (L_{T}-2 \zeta^{0})}{R}}+e^{-\frac{\pi  m (L_{T}-2 \zeta^{0})}{R}}\right)}{q^m-1},\\
  &=\sum _{m=1}^{\infty } -m^2 \text{csch}\left(\frac{\pi  L_{T} m}{2 R}\right) \sin \left(\frac{2 \pi  m \zeta^{1}}{R}\right) \cosh \left(\frac{\pi  m (L_{T}-4 \zeta^{0})}{2 R}\right),\\
  &=-i \sum _{m=1}^{\infty } \frac{ m^2 e^{-\frac{\pi  L_{T} m}{2 R}} \left(\sinh \left(2 m \varphi-\frac{2 i \pi  m \zeta^{1}}{R}\right)-\sinh \left(2 m\varphi+\frac{2 i \pi  m \zeta^{1}}{R}\right)\right)}{1-e^{-\frac{\pi  L_{T} m}{R}}},\\
\end{split}
\end{equation}
with $\varphi=\frac{\pi  (L_{T}-4 \zeta^{0})}{4 R}$.

The series can be represented as a derivative  
\begin{equation}  
 \mathcal{S}=\frac{-i R^2}{4 \pi^2} \frac{\partial^{2}}{\partial \zeta^{0} \partial \zeta^{0}} \sum _{m=1}^{\infty } \frac{e^{-\frac{\pi  L_{T} m}{2 R}} \left(\sinh \left(2 m \varphi-\frac{2 i \pi  m \zeta^{1}}{R}\right)-\sinh \left(2 m \varphi+\frac{2 i \pi  m \zeta^{1}}{R}\right)\right)}{1-e^{-\frac{\pi  L_{T} m}{R}}},
\end{equation}

The sum is encapsulated in an auxiliary function
 \begin{equation}
     \begin{split}
       U(\zeta^{0},\zeta^{1})&= \sum _{m=1}^{\infty } \frac{e^{-\frac{\pi  L_{T} m}{2 R}} \left(\sinh \left(2 m \varphi-\frac{2 i \pi  m \zeta^{1}}{R}\right)-\sinh \left(2 m \varphi+\frac{2 i \pi  m \zeta^{1}}{R}\right)\right)}{1-e^{-\frac{\pi  L_{T} m}{R}}},\\
       &=\left( \frac{\vartheta _1^{\prime }}{ \vartheta_1}\left(i\phi ,e^{-\frac{L_{T} \pi }{2 R}}\right)- \frac{\vartheta _1^{\prime }}{ \vartheta_1}\left(i\phi^{*},e^{-\frac{L_{T} \pi }{2 R}}\right)\right),
%\frac{\partial}{\partial \zeta^{1}} \sum _{m=1}^{\infty } \frac{e^{-\frac{\pi  L_{T} m}{2 R}} \left(\sinh \left(2 m \varphi-\frac{2 i \pi  m \zeta^{1}}{R}\right)}{1-e^{-\frac{\pi  L_{T} m}{R}}}d\zeta^{1}
  \label{U}
  \end{split}
\end{equation}
  such that by using Eq.\eqref{IdentityXII}- one conveniently express the series sums in terms of Jacobi function $\vartheta_{1}$.

 Assuming small value of $\zeta^{1}$ the function assumes a differential form
   \begin{equation}
     \begin{split}
       U(\zeta^{0},\zeta^{1})d\zeta^{1}=\frac{\partial}{\partial \zeta^{1}} \frac{\vartheta _1^{\prime }}{ \vartheta_1}\left(i\phi ,e^{-\frac{L_{T} \pi }{2 R}}\right)d\zeta^{1}
%\sum _{m=1}^{\infty } \frac{e^{-\frac{\pi  L_{T} m}{2 R}} \left(\sinh \left(2 m \varphi-\frac{2 i \pi  m \zeta^{1}}{R}\right)-\sinh \left(2 m \varphi+\frac{2 i \pi  m \zeta^{1}}{R}\right)\right)}{1-e^{-\frac{\pi  L_{T} m}{R}}}d\zeta^{1},\\
%       &=\frac{\partial}{\partial \zeta^{1}} \sum _{m=1}^{\infty } \frac{e^{-\frac{\pi  L_{T} m}{2 R}} \left(\sinh \left(2 m \varphi-\frac{2 i \pi  m \zeta^{1}}{R}\right)}{1-e^{-\frac{\pi  L_{T} m}{R}}}d\zeta^{1}
  \label{U}
  \end{split}
\end{equation}

    The expectation value corresponding to the correlator Eq.\eqref{W22b4} is obtained as an integral with respect to $\zeta^{0}$ over the square of the sum $\mathcal{S}$. Using Eq.\eqref{U} this reads

\begin{equation}
  W^{2}_{1,b_4}=\frac{b_4 (D-2)}{64 R^2 \sigma^2} \left\{ \lim_{\zeta^1 \to 0}+\lim_{\zeta^1 \to R} \right\} d^2\zeta^{1}\int _{\partial \Sigma} d\zeta^{0}  \left( \frac{\partial^{2}}{\partial \zeta^{0^{2}}} U(\zeta^{0},\zeta^{1})  \right)^{2}.
\label{W22b4}  
\end{equation}

   The square of the derivative of $U(\zeta^{0},\zeta^1)$ is manipulated through integrating two times by parts

\begin{equation}
\begin{split}
\int_{0}^{L_{T}} \left(\frac{\partial^2 U}{\partial \zeta^{0^2}}\right)^2 \, d\zeta^{0}= \Bigg[\frac{\partial U}{\partial \zeta^{0}}\frac{\partial^2 U}{\partial \zeta^{0^2}} \,- U  \frac{\partial^3 U}{\partial \zeta^{0^3}} \Bigg]^{L_{T}}_{0}+ \int_{0}^{L_{T}} U \frac{\partial^4 U}{\partial \zeta^{0^4}} d\zeta^{0}\, 
\end{split}
\label{ddU1}
\end{equation}
Then expanding $U(\zeta^{0},\zeta^{1})$ in the integrand of the last term in Eq.~\eqref{ddU1} around $\frac{\pi L_{T}}{R}$  
\begin{equation}
\begin{split}
  \int_{0}^{L_{T}} \left(\frac{\partial^2 U}{\partial \zeta^{0^2}}\right)^2 \, d\zeta^{0}=& \Bigg[\frac{\partial U}{\partial \zeta^{0}}\frac{\partial^2 U}{\partial \zeta^{0^2}} \,- U  \frac{\partial^3 U}{\partial \zeta^{0^3}} \Bigg]^{L_{T}}_{0}+2U\left(\frac{L_{T}}{4R},\zeta^{1}\right)\frac{\partial^3 U}{\partial \zeta^{0^3}}\Bigg|^{L_{T}/2}_{0}+2 \sum_{n=1}^{\infty }  \left(\frac{i\pi}{R}\right)^{n} \\
  &\times U^{(n)}\left(\frac{L_{T}}{4},\zeta^{1}\right)\int_0^{L_{T}/2} \left(\zeta^{0}-\frac{L_{T}}{4}\right)^{n} \frac{\partial^4 U}{\partial \zeta^{0^4}} d\zeta^{0}\, 
\end{split}
\label{ddU2}
\end{equation}

 Substituting Eq.\eqref{U} into Eq.\eqref{ddU2} and Eq.\eqref{W22b4}, the correlator becomes   
%  Differentiating the last term in the above integral with respect to $\zeta^{0}$, 
%\begin{equation}
%\begin{split}
%  \mathcal{S}_{3} &=\frac{\partial }{\partial \zeta^{0}}\frac{\partial }{\partial \zeta^{0}} U(\zeta^0,\zeta^1)\\
%&=\frac{4 i \pi^2}{R^2} \sum _{m=1}^{\infty }\frac{m^3 e^{-\frac{\pi  L_{T} m}{2 R}}}{1-e^{-\frac{\pi  L_{T} m}{R}}}\\
%&~~~~\times \left(\sinh \left(\frac{\pi  m (L_{T}-4 \zeta^{0})}{2 R}-\frac{2 i \pi  m \zeta^{1}}{R}\right)-\sinh \left(\frac{\pi  m (L_{T}-4 \zeta^{0})}{2 R}+\frac{2 i \pi  m \zeta^{1}}{R}\right)\right),
%\end{split}
%\end{equation}
%  Expanding the denumerator we found that, 
%\begin{equation}
% \begin{split} 
%   \mathcal{S}_{3} &=\frac{4 i \pi^2}{R^2} \sum _{k=0}^{\infty } \sum _{m=1}^{\infty } m^3 e^{-\frac{(2 k+1) (\pi  L_{T} m)}{2 R}}\\ 
%   &\times \Bigg(\sinh \left(\frac{\pi  m (L_{T}-4 \zeta^{0})}{2 R}-\frac{2 i \pi  m \zeta^{1}}{R}\right)  -\sinh \left(\frac{\pi  m (L_{T}-4 \zeta^{0})}{2 R}+\frac{2 i \pi  m \zeta^{1}}{R}\right)\Bigg).\\
%   &=0.
% \end{split}
% \label{S3}
%\end{equation}
%  since series sum Eq.\eqref{S3} assumes the same form as Eq.\eqref{S1plusS2} which yields a trivial contribution.
\begin{equation}
 \begin{split}
   W^{2}_{1,b_4}& = \frac{b_4 \pi^{3}(D-2)}{32 R^5 \sigma^2}\Bigg[ \frac{\vartheta^{\prime}_{1}}{\vartheta_{1}}\left(0,q\right)^{(2)} \frac{\vartheta^{\prime}_{1}}{\vartheta_{1}}\left(0,q\right)^{(3)}-\frac{\vartheta^{\prime}_{1}}{\vartheta_{1}}\left(0,q\right)^{(1)} \frac{\vartheta^{\prime}_{1}}{\vartheta_{1}}\left(0,q\right)^{(4)}+\frac{\vartheta^{\prime}_{1}}{\vartheta_{1}}\left(\frac{\pi \tau}{2},q\right)^{(1)}\\
     &\times 2\frac{\vartheta^{\prime}_{1}}{\vartheta_{1}}\left(0,q\right)^{(4)}+...\Bigg],
  \end{split}
\label{BL}
\end{equation}

%  Owing to the smallness of the the derivatives $U^{(n)}\left(\frac{L_{T}}{4},\zeta^{1}\right)\simeq 0$, the remainder of the expansion $\sum_{n=1}^{\infty }\epsilon=\zeta(\infty)$ of the last term in the above Eq.\eqref{BL} is zeta regularized to 1.
  
  The remnant divergences haunting in the n-th derivative of the logarithmic derivative at the pole $\zeta=0$ are again eliminated with $\mathbf{\zeta}$-function Eq.\eqref{IdentityVI:eq3} and Eq.\eqref{IdentityVI:eq4}. The third partial derivative outturns to 
\begin{equation}
\begin{split}   
\frac{\vartheta^{\prime}_{1}}{\vartheta_{1}}\left(0,q\right)^{(3)}&=32\sum_{n=1}^{\infty}\frac{n^3 q^{2n}}{1-q^{2n}}- 16\zeta(-3),\\
&=   \frac{32}{240}\big(E_{4}(2\tau)-1\big)-\frac{16}{120}.
\end{split}  
\label{Reg1b4}
\end{equation}
 Similarly the fourth partial derivative is
\begin{equation}
\begin{split}   
\frac{\vartheta^{\prime}_{1}}{\vartheta_{1}}\left(0,q\right)^{(4)}&=\sum_{n=1}^{\infty}\epsilon-32 i\sum_{n=1}^{\infty}n^4,\\
&= \zeta(\infty)+i\zeta(-4).\\
&=1\quad.
\end{split}
\label{Reg2b4}
\end{equation}
% The expectation value finally turn out to be
  The expectation value Eq.\eqref{BL}, after parameter redefinition $b_2 ~d^2\zeta^{1} \longrightarrow b_2$ and making use of the regularization Eq.~\eqref{Reg1b4} and Eq.~\eqref{Reg2b4}, eventually be

\begin{equation}
\begin{split}
  W^{2}_{1,b_4}
 % =&\frac{\pi^3 (D-2)b_4 }{32 R^5 \sigma^2} \Bigg[\Bigg(\left( \frac{32}{240}\left(E_{4}(2\tau)-1\right)-\frac{16}{120} \right)-\left( \frac{8}{24}\left(1-E_{2}(2\tau)\right)+\frac{1}{12}\right)+h\left(\frac{\tau}{2} \right)\Bigg)\Bigg]\\
             =&-\frac{\pi^3 (D-2)b_4 }{32 R^5 \sigma^2} \Bigg(  \frac{2}{15}E_{4}\left(2\tau \right)+\frac{1}{3}E_{2}\left(2\tau \right)-\frac{7}{60}-2\frac{\vartheta _1^{\prime }}{\vartheta _1}\left(\frac{i\pi }{2}\tau,q\right)^{(1)} \Bigg).
\end{split}
\end{equation}

%  \left( \frac{22}{3}\text{E}_2(\tau/2)-\frac{40}{3} \text{E}_2(\tau)-110 \right)% \left(61+4 \,\text{E}_2(\tau)-36 \,\text{E}_2(2\tau)\right).
\subsection{The expectation value of $W^{2}_{2,b_4}$}~\\
  The second Wick contraction of \eqref{W2b4total} is
\begin{equation}
  W_{2,b_4}^{2}=\langle \left(\mathbf{X} \cdot \mathbf{X}\right) \partial_{0}^2\partial_{1}  \partial_{0}^2\partial_{1} \left(\mathbf{X} \cdot \mathbf{X}\right) \rangle 
\end{equation}

\noindent The point-split correlator in terms of Green functions reads
\begin{equation}
  W^{2}_{2,b_4}=\int _{\partial \Sigma} d\zeta^0  G\left(\zeta^0,\zeta^1;\zeta^{0'},\zeta^{1'}\right)\partial _{0}^2\partial_{1}\partial _{0'}^2\partial_{1'} G\left(\zeta^0,\zeta^1;\zeta^{0'},\zeta^{1'}\right). 
  \label{W21b2}
\end{equation}
Substituting the free propagator Eq.\eqref{GreenPropagator} into the above Eq.\eqref{W21b2}, the expectation value is
\begin{equation}
\begin{split}
  W^{2}_{2,b_4}&=\int _{\partial \Sigma}  d\zeta^0 \Bigg(\frac{\pi ^4 m^5}{n R^6 \sigma^2 \left(q^m-1\right) \left(q^n-1\right)} e^{-\frac{\pi  (m+n) \left(\zeta^{0}-\zeta^{0'}\right)}{R}} \left(q^m e^{\frac{2 \pi  m \left(\zeta^{0}-\zeta^{0'}\right)}{R}}+1\right)  \\
  & \left(q^n e^{\frac{2 \pi  n \left(\zeta^{0}-\zeta^{0'}\right)}{R}}+1\right) \cos \left(\frac{\pi  m \zeta^{1'}}{R}\right) \cos \left(\frac{\pi  m \zeta^{1}}{R}\right) \sin \left(\frac{\pi  n \zeta}{R}\right) \sin \left(\frac{\pi  n \zeta^{1'}}{R}\right)\Bigg).
  \label{W1b4}
\end{split}
\end{equation}
the point-splitting implies the infinitesimal transform 
\begin{equation}
  \begin{split}
\zeta^{1'}=\zeta^{1}+\epsilon,\\
\zeta^{0'}=L_{T}-\zeta^{0}-\epsilon'.\\
\end{split}
\end{equation}
   Substituting the transform into Eq.\eqref{W1b4} and considering the limits $\epsilon\rightarrow 0$ and $\epsilon' \rightarrow 0$, the expectation value becomes
\begin{equation}
  \begin{split}
  W^{2}_{2,b_4}&=\dfrac{\pi^4}{R^6 \sigma^2}\int _{\partial \Sigma}  d\zeta^0 \Bigg(\frac{ m^5}{n  \left(q^m-1\right) \left(q^n-1\right)} e^{-\frac{2 \pi  m (L_{T}-2 \zeta^{0})}{R}}  \left(e^{\frac{2 \pi  m (L_{T}-2 \zeta^{0})}{R}}+q^m\right)\\
  & \left(q^n e^{\frac{\pi(m-n)(L_{T}-2 \zeta^{0})}{R}} +e^{\frac{\pi (m+n) (L_{T}-2 \zeta^{0}) }{R}}\right)
  \cos ^2\left(\frac{\pi  m \zeta^{1}}{R}\right) \sin ^2\left(\frac{\pi  n \zeta^{1}}{R}\right)\Bigg).
  \label{W21b4}
  \end{split}
\end{equation}
  Consider each sum over the separate indicies in the above integral, the first series 
\begin{equation}
  \begin{split}
    \mathcal{S}_1&=\sum_{m=1}^{\infty}\frac{1}{q^{2m}-1} \left(m^5 e^{-\frac{2 m \pi(L_{T}-2 \zeta^{0})}{R}} e^{\frac{m \pi (L_{T}-2 \zeta^{0})}{R}} \cos ^2\left(\frac{m \pi \zeta^{1}}{R}\right) q^{2m}+e^{\frac{2 \pi  m (L_{T}-2 \zeta^{0})}{R}}\right),\\
                &=m^5 \text{csch}\left(\frac{\pi m L_{T} }{2 R}\right) \cos ^2\left(\frac{\pi  m \zeta^{1}}{R}\right) \cosh \left(\frac{\pi  m (3 L_{T}-4 \zeta^{0})}{2 R}\right),
    \end{split}
\end{equation}
   have similar functional form to Eq.\eqref{S1sum} with the replacement  $m^{5} \rightarrow m^3$ in Eq.\eqref{S11} and Eq.\eqref{S12}. Similar $\mathbf{\zeta}$ regularization procedures allows to reduce the series into the  
\begin{equation}
\begin{split}
\mathcal{S}_1=& \sum _{k=0}^{\infty } \sum _{m=1}^{\infty } (\mathcal{S}_{1,1}(m^5,k)+\mathcal{S}_{1,2}(m^5,k)),\\
=& \zeta(\infty),\\
=& 1.
\end{split}
\end{equation}
The second series 
\begin{equation}
  \mathcal{S}_2= \sum_{n=1}^{\infty} \frac{1}{n} \text{csch}\left(\frac{\pi  nL_{T}}{2R}\right) \sin ^2\left(\frac{\pi n\zeta^{1}}{R}\right) \cosh \left(\frac{\pi n(3 L_{T}-4\zeta^{0})}{2R}\right),
\label{S24}
\end{equation}
  assumes the same form as Eq.\eqref{S2}. 
  %+\frac{\pi^3 (D-2)b_4}{128 R^5 \sigma^2} \left(\frac{4}{3}E_2(2\tau)-\frac{5}{3}\right) \left(61+4 \text{E}_2(\tau)-36 \text{E}_2(2\tau)\right)\epsilon^2

  \begin{equation}
  \begin{split}
    W^{2}_{2,b_2}& =-\frac{\pi^{6} b_4 (D-2)}{ R^7 \sigma^2}. \\
%    \left\{ \lim_{\zeta^1 \to 0}+\lim_{\zeta^1 \to R} \right\} \sum _{k=1}^{\infty } \frac{e^{-\frac{k (\pi  L)}{R}}  \sinh \left(\frac{2 \pi  k L}{R}\right) \sin ^2\left(\frac{\pi  k \zeta^{1}}{R}\right)}{  k^2 \left(1-e^{-\frac{k (\pi  L)}{R}}\right)}
\end{split}
\label{I11}
\end{equation}
% \left(\frac{\vartheta _1^{\prime}\left(i\tau/2,e^{-\frac{L_{T} \pi }{2 R}}\right)}{\vartheta _1\left(\frac{i \pi  \zeta^{0}}{2 R},e^{-\frac{L_{T} \pi }{2 R}}\right)}-2\right)
%%%%%%%%%%%%%%%%%%%%%%%%%%%%%%%%%%%%%%
%%%%%%%%%%%%%%%%%%%%%%%%%%%%%%%%%%%%%%
%\bibliography{Biblio2.bib}

\end{document}